\documentclass[superscriptaddress,reprint,showpacs,nofootinbib,amsmath,amssymb,aps,prstab]{revtex4-1}

\usepackage{graphicx}
\usepackage{dcolumn}
\usepackage{bm}
\usepackage{siunitx}
\usepackage{upgreek}
\usepackage{color}
\usepackage{hyperref}

\begin{document}

\title{Vacancy-Hydrogen Interaction in Niobium during Low-Temperature Baking} 

\author{Marc Wenskat}
\email{marc.wenskat@desy.de}
\affiliation{Institute of Experimental Physics, University of Hamburg, Luruper Chaussee 149, 22761 Hamburg Hamburg, Germany}
\affiliation{Deutsches Elektronen-Synchrotron, Notkestrasse 85, 22607 Hamburg, Germany} 
\author{Jakub \v{C}i\v{z}ek}
\affiliation{Faculty of Mathematics and Physics, Charles University, V Hole\v{s}ovi\v{c}kach 2, 180 00 Praha 8, Czech Republic}
\author{Maciej Oskar Liedke}
\affiliation{Institute of Radiation Physics, Helmholtz-Zentrum Dresden-Rossendorf, Bautzner Landstr. 400, 01328 Dresden, Germany}
\author{Maik Butterling}
\affiliation{Institute of Radiation Physics, Helmholtz-Zentrum Dresden-Rossendorf, Bautzner Landstr. 400, 01328 Dresden, Germany}
\author{Christopher Bate}
\affiliation{Institute of Experimental Physics, University of Hamburg, Luruper Chaussee 149, 22761 Hamburg Hamburg, Germany}
\affiliation{Deutsches Elektronen-Synchrotron, Notkestrasse 85, 22607 Hamburg, Germany} 
\author{Petr Hau\v{s}ild}
\affiliation{Faculty of Nuclear Sciences and Physical Engineering, Czech Technical University in Prague, Trojanova 13, 120 01 Praha 2, Czech Republic} 
\author{Eric Hirschmann}
\affiliation{Institute of Radiation Physics, Helmholtz-Zentrum Dresden-Rossendorf, Bautzner Landstr. 400, 01328 Dresden, Germany} 
\author{Andreas Wagner}
\affiliation{Institute of Radiation Physics, Helmholtz-Zentrum Dresden-Rossendorf, Bautzner Landstr. 400, 01328 Dresden, Germany} 
\author{Hans Weise}
\affiliation{Deutsches Elektronen-Synchrotron, Notkestrasse 85, 22607 Hamburg, Germany} 

\date{\today} 

\begin{abstract}
A recently discovered modified low-temperature baking leads to reduced surface losses and an increase of the accelerating gradient of superconducting TESLA shape cavities. We will show that the dynamics of vacancy-hydrogen complexes at low-temperature baking lead to a suppression of lossy nanohydrides at 2\,K and thus a significant enhancement of accelerator performance. Utilizing Doppler broadening Positron Annihilation Spectroscopy, Positron Annihilation Lifetime Spectroscopy and instrumented nanoindentation, samples made from European XFEL niobium sheets were investigated. We studied the evolution of vacancies in bulk samples and in the sub-surface region and their interaction with hydrogen at different temperature levels during {\it in-situ} and {\it ex-situ} annealing.

\end{abstract}


\maketitle

\section{\label{sec:level1}Introduction}
The influence of hydrogen on the rf losses of cavities and the necessity of outgassing has been known for quite some time. The so-called 'hydrogen Q-disease' marks an increase of the surface losses at cryogenic temperatures which sets in even at low values of the applied accelerating field  \cite{Bonin1991, Antoine2003} and it works as follows: the operating temperature of superconducting accelerating cavities is 2-4\,K, and while crossing the temperature range of 200-50\,K during cool down, different phases of niobium hydrides form on the surface and in the rf penetration layer, causing increased losses as niobium hydrides are only superconducting below 1.3\,K \cite{Isagawa1980}. To prevent this Q-disease from happening, cavities need to be annealed at $800-900^\mathrm{o}\,\mathrm{C}$ for 3h in a vacuum furnace at pressures below $10^{-5} \, \mathrm{mbar}$ to purify the material (by hydrogen removal), hence preventing the formation of hydrides during cool down. In particular, the surface underwent chemical treatments to improve the surface roughness causing a significant hydrogen uptake \cite{Aderhold2010,Ciovati2010a}.
\newline  
Open volume lattice defects (vacancies, vacancy clusters, dislocations), however, are known to have high trapping potential for interstitial impurities, especially hydrogen \cite{Cottrell1949, Chemical2006, Cizek2009, Romanenko2010, Romanenko2013, Ford2013b}, and after the high temperature bake a fraction of several hundreds ppm of hydrogen remains in the lattice \cite{Ciovati2010a} in the near surface layer.  The formation of so-called “nanohydrides” \cite{DeGennes1965, Romanenko2013}, which are only weakly superconducting by proximity effect \cite{Fauchere1997} below a certain threshold of an applied accelerating field, causes losses above this threshold. The threshold value for the high-field Q-slope is 90-100\,mT, which relates to a maximum niobium hydride size of roughly 10\,nm\cite{DeGennes1965}. This mechanism is proposed as the origin of the so-called 'high-field Q-slope', describing an increase of surface resistance above an onset field, which can be overcome by a $120^\mathrm{o}\,\mathrm{C}$ bake for 48\,h in a pressure below $10^{-6} \, \mathrm{mbar}$ which is an empirical cure \cite{Visentin2010, Ciovati2007a, Ciovati2010a, Romanenko2013b, Trenikhina2015c}. 
\newline
A recent unintended discovery has modified the $120^\mathrm{o}\,\mathrm{C}$ baking procedure by starting with a unintended intermediate step of $75^\mathrm{o}\,\mathrm{C}$ for 4\,h before going to the final temperature for the remaining 44\,h. This modified bake, meanwhile called 'low-T bake', showed a reduction of the observed losses by a factor of two and increased the achievable accelerating field by 10-20\% in comparison to the standard treatment according to \cite{Grassellino2018}. Literature research identified the $70^\mathrm{o}\,\mathrm{C}$ as a specific temperature, at which the arrangement of vacancies and dislocations starts to change but most of the interstitials, except hydrogen, in the lattice and the oxides on the surface are still immobile \cite{Stanley1967}. In addition, according to the Nb-H phase diagram at $74^\mathrm{o}\,\mathrm{C}$ a phase transition between $\alpha'$- and $\beta$- NbH takes place \cite{Gupta1994}. 
\newline
It is plausible that the new baking procedure might influence the concentration of vacancies and their interaction with hydrogen in the relevant rf penetration layer, i.e., sub-surface region of about 100-200\,nm, in a beneficial way to prevent formation of lossy nanohydrides as v+nH complexes may serve as nucleation sites as has been suggested in \cite{Barkov2012, Barkov2013,Romanenko2013b}\footnote{It is known that interstitial nitrogen inside the niobium lattice traps hydrogen as efficiently as vacancies but might prevent the formation of hydrides \cite{Richter1976, Pfeiffer1976}, which might be part of the reason for the improvement of cavities as described in \cite{Grassellino2013a, Grassellino2017d, Sung2019}}. 
\newline
Up to now, no fundamental theory exists linking surface properties such as roughness, chemical composition, grain structure and grain orientation and rf properties of cavities. Re-assessing the assumed processes happening during standard treatment of cavities utilizing modern surface analysis techniques might help to find a common link and improve our fundamental understanding of superconducting radio-frequency (srf) cavities.
In addition, cost-saving effort for superconducting accelerators are always highly appreciated, especially with planned larger upgrades or construction of research facilities like the European XFEL, LCLS-II and SHINE \cite{Brinkmann2014a,Michizono2017, Raubenheimer2018a, Zhu2018}. Potentially saving substantial fractions of the planned investments is highly desirable. 
\newline
In this manuscript we will show that positron annihilation spectroscopy serves as a dedicated tool to clarify several open issues: (i) vacancy kinetics, the threshold temperature and baking duration at which different processes start. (ii) The understanding of vacancies and interstitial hydrogen interactions and the formation of complexes in which hydrogen atoms are associated with vacancies \cite{Fukai1994} at these temperatures is still missing. And finally (iii) a comparison of material from different niobium vendors for cavities and how this might affect the performance is of large importance.

\section{\label{sec:level2}Methods}
\subsection{\label{sec:level2.1}Samples}
\label{sec:sample_prep}

The samples used for this study were cut out of niobium sheets from the European XFEL cavity production. The sheets were produced by Tokyo Denkai Co. Ltd. ('Tokyo Denkai' - samples with numbers above \#50) or Ningxia Orient Tantalum Industry Co. Ltd. ('Ningxia' - samples with numbers below \#50). Mechanical and chemical properties of Nb sheets supplied by the both producers are summarized in Table \ref{tab:samplespec}. 
\begin{table}[htbp]
\caption{\label{tab:samplespec} Material properties of the niobium sheets: Vickers hardness (HV), tensile strength (R\textsubscript{m}), yield strength (R\textsubscript{p}), concentration of oxygen (c\textsubscript{O}), nitrogen (c\textsubscript{N}), hydrogen (c\textsubscript{H}) and carbon (c\textsubscript{C}).}
		\begin{tabular}{l c c }
 			 &  \textbf{\normalsize{Tokyo Denkai}}    & \textbf{\normalsize{Ningxia}} \\   
\hline		
		{\normalsize{HV}}     &	42.5-50.6 			& 37-46  \\
		{\normalsize{$\mathrm{R_m [N/mm^2]}$}}     &	170 			& 160  \\
		{\normalsize{$\mathrm{R_p [N/mm^2]}$}}     &	62 			& 62  \\
		{\normalsize{$\mathrm{c_O [ppm\,at]}$}}     &	3.5-4.5 			& 5  \\
		{\normalsize{$\mathrm{c_N [ppm\,at]}$}}     &	3.2-4.4 			& 5  \\
		{\normalsize{$\mathrm{c_H [ppm\,at]}$}}     &	0.5-0.8 			& 1  \\
		{\normalsize{$\mathrm{c_C [ppm\,at]}$}}     &	0.2-0.7 			& 6  \\
		\hline
		\end{tabular}
\end{table}
The samples are of flat-conical design, with a base diameter of 12\,mm, a top diameter of 10\,mm and a thickness of 2.8\,mm. They were cut from the sheets by a water jet cutter and turned to their final shape. 
The sample preparation followed closely the standard cavity preparation and the intermediate cleaning and rinsing was carried out in an ISO 4 cleanroom environment \cite{Altarelli2007,Aderhold2010, Singer2016, Reschke2017}:
\begin{enumerate}
	\item Electro-polishing (EP) with a total removal of $120\, \upmu \mathrm{m}$ surface layer
	\item Ultra-sonic Cleaning and High Pressure Rinsing (DI Water with 100\,bar)
	\item $800^\mathrm{o}\,\mathrm{C}$ annealing for 3 hours at a pressure below $10^{-5} \mathrm{mbar}$
	\item Electro-polishing with a total removal of $30\, \upmu \mathrm{m}$ surface layer
	\item Ultra-sonic Cleaning and High Pressure Rinsing (DI Water with 100\,bar)
\end{enumerate}
The electro-poslishing was done with HF (w=40\%) and H\textsubscript{2}SO\textsubscript{4} (w=98\%) acids mixed in the volume ratio 1:9. Dedicated sample chemistry and high pressure rinsing holders have been developed to use the standard infrastructure for cavity treatment at DESY. The final preparation step, the low temperature bake, was then studied with {\it in-situ} and {\it ex-situ} approaches using different positron annihilation spectroscopy setups. An overview of the samples and the measurements is given in Table \ref{tab:SamplePrep}. A strength of our approach is that samples from the same sheets were studied with complementary methods which further strengthened the interpretation of each individual result.   

\begin{table*}[htbp]
\caption{\label{tab:SamplePrep} List of samples used, including material and temperature treatment applied. For details of the used methods see section \ref{sec:level2.2}. }
\begin{ruledtabular}
		\begin{tabular}{c l l l l }
 			 \textbf{\normalsize{Sample}}    &  \textbf{\normalsize{Material}}    & \textbf{\normalsize{Method}} &  \textbf{\normalsize{Facility}} &  \textbf{\normalsize{Treatment}} \\   
\hline		
		\textbf{\normalsize{6}}     &	Ningxia 			& PALS, CDB  & Prague & $70^\mathrm{o}\,\mathrm{C}$ for 4\,h, $120^\mathrm{o}\,\mathrm{C}$  for 4, 40, 180\,h in $\mathrm{p}\leq 10^{-3}\,\mathrm{mbar}$   \\ 
		\textbf{\normalsize{73}}    &	Tokyo Denkai 	& PALS, CDB   & Prague & $70^\mathrm{o}\,\mathrm{C}$ for 4\,h, $120^\mathrm{o}\,\mathrm{C}$  for 4, 40, 180\,h in $\mathrm{p}\leq 10^{-3}\,\mathrm{mbar}$   \\ 
		
		\textbf{\normalsize{14}}    &	Ningxia  			& DB-VEPAS 			  & AIDA 	& DESY sample furnace at $70^\mathrm{o}\,\mathrm{C}$ for 4\,h in $\mathrm{p} \approx 10^{-6}\,\mathrm{mbar}$ \\ 
		\textbf{\normalsize{17}}    &	Ningxia  			& DB-VEPAS 			  & AIDA 	& Steps from $70^\mathrm{o}\,\mathrm{C}$ to $120^\mathrm{o}\,\mathrm{C}$ for 4\,h each in $\mathrm{p} \approx 10^{-10}\,\mathrm{mbar}$  \\ 
		\textbf{\normalsize{64}}    &	Tokyo Denkai  & DB-VEPAS 			  & AIDA 	& Steps from $70^\mathrm{o}\,\mathrm{C}$ to $120^\mathrm{o}\,\mathrm{C}$ for 4\,h each in $\mathrm{p} \approx 10^{-10}\,\mathrm{mbar}$ \\ 
		\textbf{\normalsize{78}}    &	Tokyo Denkai  & VEPALS				  & MePS 	& Steps from $70^\mathrm{o}\,\mathrm{C}$ to $120^\mathrm{o}\,\mathrm{C}$ for 4\,h each in $\mathrm{p}\leq 10^{-7}\,\mathrm{mbar}$ \\ 
		\hline
		\end{tabular}
		\end{ruledtabular}
\end{table*}
\subsection{Instrumented Nanoindentation}
Instrumented nanoindentation measurements were performed on Anton Paar NHT2 Nanoindentation Tester with Berkovich diamond indenter. Continuous multi-cycle indentations with increasing load were performed varying the maximum load at each cycle from 1\,mN up to 500\,mN. This type of load cycle allows characterization at different depth levels in the same position, see e.g. \cite{Hausild2016, Hausild2018}. Loading time was 10\,s per each loading cycle followed by 5\,s hold at the maximum load and unloading time of 10\,s for each cycle. The results (indentation hardness) were evaluated using the Oliver-Pharr method according to the ISO\,14577 standard. 

\subsection{\label{sec:level2.2}Positron Annihilation Spectroscopy}
\begin{figure}[htpb!]
	\centering
		\includegraphics[width=1.0\columnwidth]{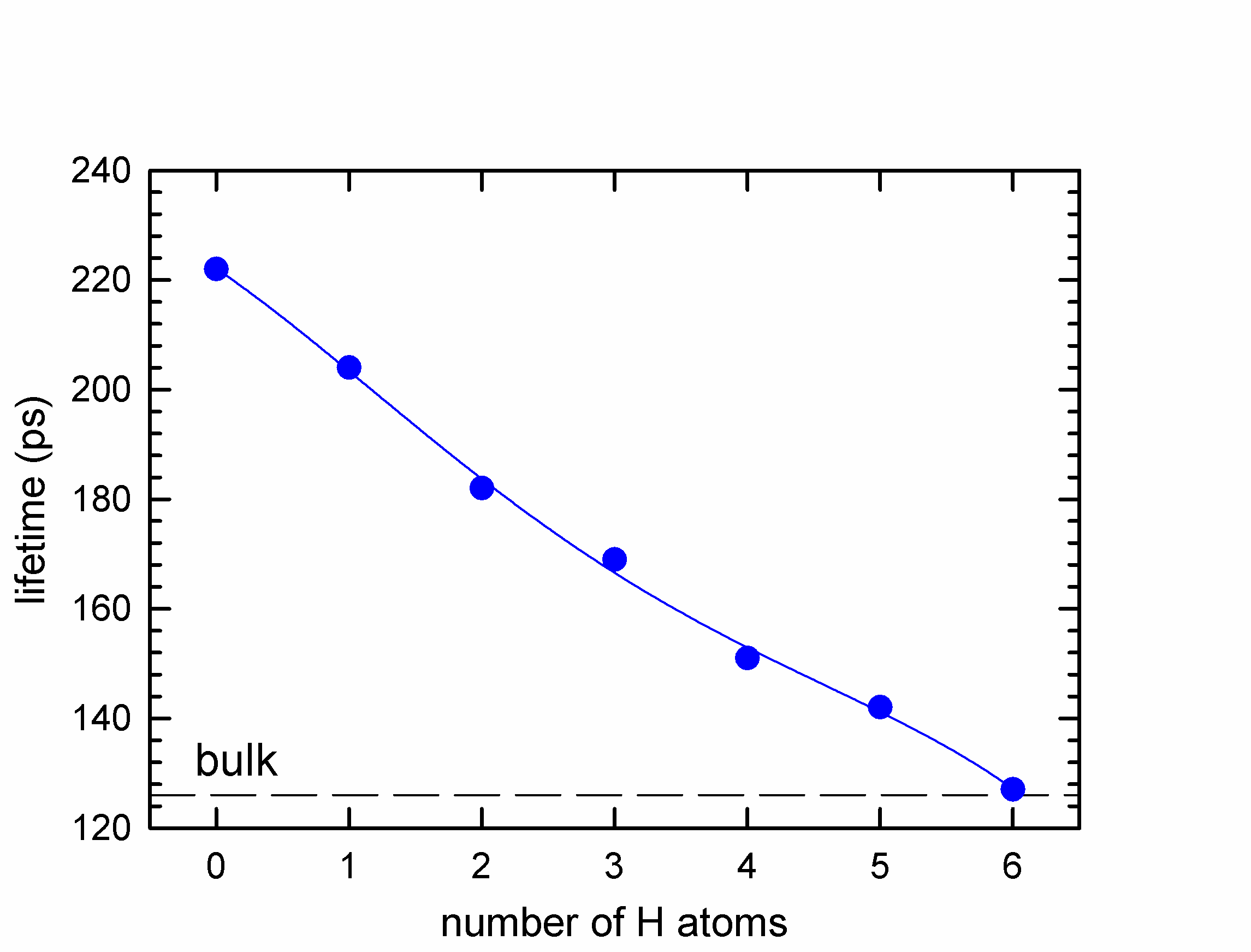}
	\caption{Dependence of lifetime $\tau_{v+H}$ of positron
trapped at vacancy-hydrogen complex on the number of hydrogen atoms surrounding the vacancy calculated with ab-initio methods \cite{Cizek2009}.}
	\label{fig:lifetime_vnh}
\end{figure}

As positrons can easily get trapped in metal vacancy defects and are known to be very sensitive to their chemical environment, they are the optimal choice to analyze the types, concentrations and atomic environment of vacancies \cite{Cizek2018}. In principle two approaches can be followed: (i) positron annihilation lifetime spectroscopy (PALS) which is based on measurement of positron lifetime in the material; (ii) Doppler broadening (DB) of the annihilation photo-peak, which is based on measurement of the Doppler shift of annihilation radiation \cite{Krause-Rehberg1999}. 
The positron lifetime is governed by local electron density at the positron annihilation site, i.e. PALS  provides information about the type and concentration of lattice defects in a material, e.g. see Fig \ref{fig:lifetime_vnh}. 

The concentration of vacancies in the samples studied was calculated using two state trapping model \cite{West1973}. The model is based on the following assumptions: (i) the sample contains single type of defects; (ii) only thermalized positrons are trapped at defects; (iii) no de-trapping of positrons already confined in defects takes place; (iv) defects are distributed uniformly and positron trapping is capture limited, i.e. it is governed by the quantum transition from the delocalized state to the trapped state in defect. 
\\
These assumptions are fulfilled in the Nb samples containing vacancy-hydrogen complexes. Positron can be either delocalized in the free state or trapped in a vacancy. The probability that at certain time $t$ (measured from the time of positron implantation into the sample) a positron is alive in the free state or in the trapped state in vacancy is denoted $n_f(t)$ and $n_v(t)$, respectively. Assuming that assumptions (i)-(iv) hold the probabilities$n_f(t)$, $n_v(t)$ are governed by the set of differential equations 
\begin{equation}  \label{eq:1}
    \frac{dn_f(t)}{dt}=-\lambda_B n_f(t)-K_v n_f(t)
\end{equation}
\begin{equation}  \label{eq:2}
    \frac{dn_v(t)}{dt}=-\lambda_v n_v(t)+K_v n_v(t)
\end{equation}
where $\lambda_B$ and $\lambda_v$ stand for the annihilation rate of positron in the free and in the trapped state in vacancy, respectively. The positron trapping rate $K_v$ is directly proportional to the concentration of vacancies $K_v=\nu_v c_v$, where $\nu_v  = 10^{14}s^{-1}$ is the specific positron trapping rate to vacancy \cite{Krause-Rehberg1999}. 
Solving the set of differential equations \ref{eq:1} and \ref{eq:2} with the initial conditions $n_f(t=0)=1$ and $n_v (t=0)=0$ results in the sum of two exponential components 
\begin{equation} \label{eq:3}
   n(t)=n_f (t)+n_v (t)= I_1 e^{-(\lambda_B+K_v)t}+I_2 e^{-\lambda_v t}
\end{equation}
where $I_1$, $I_2$ is the relative intensity of the free positron component and the contribution of positrons trapped in vacancies
and $I_1+I_2=100\%$. Since the positron lifetime spectrum is  $S(t)=-\frac{dn(t)}{dt}$ it can be expressed within the two state trapping model as
\begin{equation} \label{eq:4}
    S(t)=\frac{I_1}{\tau_1}e^{-\frac{t}{\tau_1}}+\frac{I_2}{\tau_2}e^{-\frac{t}{\tau_2}}
\end{equation}
This model function convoluted with the resolution function of corresponding spectrometer was fitted to the experimental data. From comparison of Eqs. \ref{eq:3} and \ref{eq:4} one obtains that lifetime of the free positron component is $\tau_1=\frac{1}{\lambda_B+K_v}$  and lifetime of the second component is $\tau_2=\frac{1}{\lambda_v}$. The positron trapping rate to defects is given by the equation $K_v=\frac{I_2}{I_1}  (\lambda_B-\lambda_v )$. Hence the concentration of vacancies can be calculated within the two-state simple trapping model as  
\begin{equation}
    c_v=\frac{1}{\nu_v}   \frac{I_2}{I_1}  (\frac{1}{\tau_B} -\frac{1}{\tau_2}).
    \label{cv}
\end{equation} 
The positron bulk lifetime (i.e. lifetime of positrons in a perfect, defect-free Nb lattice) for Nb is $\tau_B = \frac{1}{\lambda_B}$ = 125\,ps. \cite{Cizek2004} 
\newline
The Doppler shift of the annihilation radiation is determined by the momentum of the electron which annihilated the positron. In the DB spectroscopy, the broadening of the annihilation photo-peak is characterized in terms of the line shape parameters S (sharpness) and W (wing) which contain information about the contributions of annihilations by low or high momentum electrons, respectively.
\begin{figure}[htpb!]
	\centering
		\includegraphics[width=1.0\columnwidth]{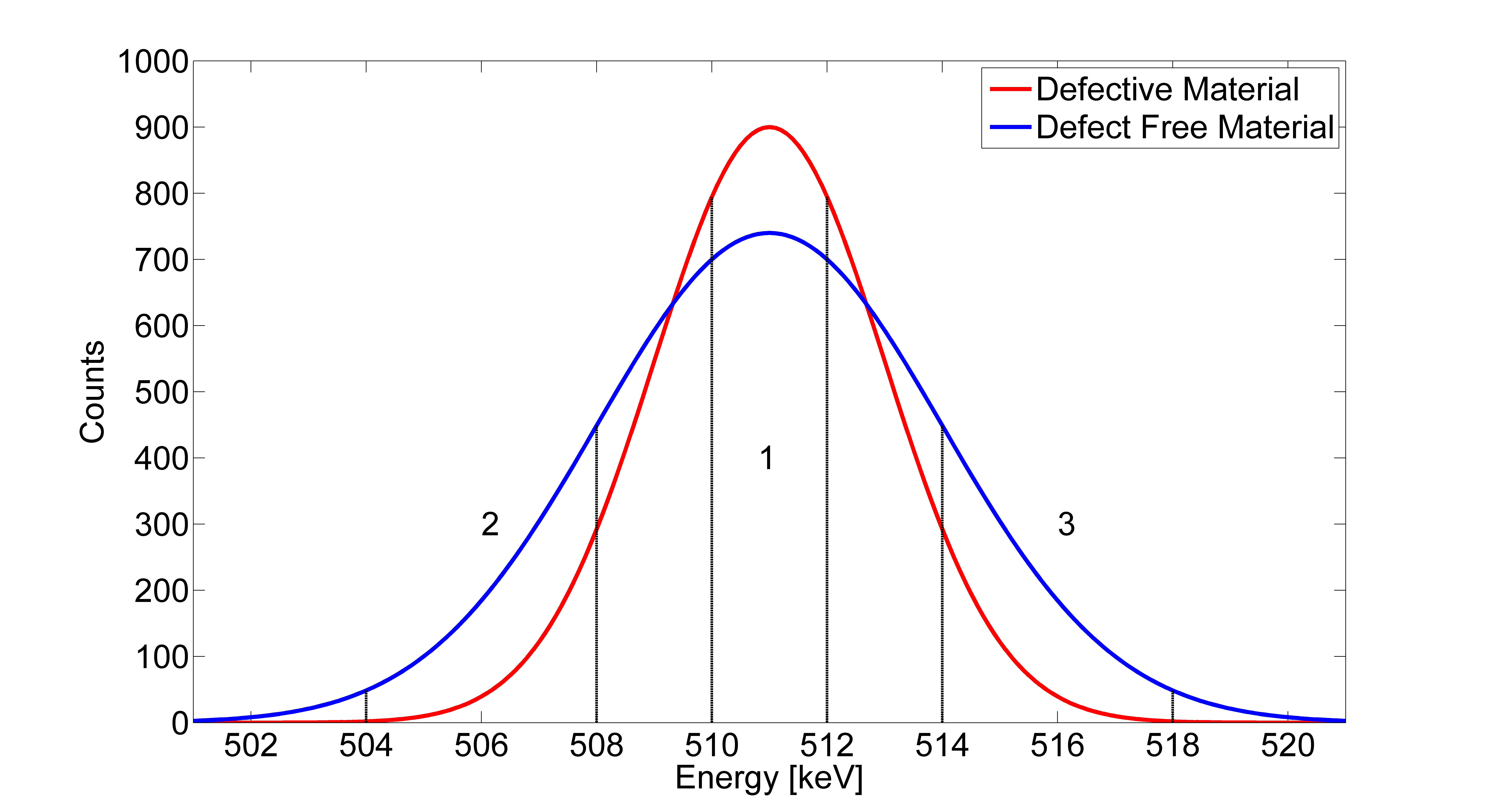}
	\caption{Positron annihilation photo-peak in a metallic material. The central area $A_1$is used to quantify the S-parameter, while the areas $A_2$ and $A_3$ are used for the W-parameter. Changes in the density, type and chemical environment of positron traps change the momentum distribution of electrons annihilating positrons and hence lead to different values of the S and W parameters.}
	\label{fig:Definition_SW}
\end{figure}
The S parameter is defined as the ratio of the central area $\mathrm{A_1}$ of the annihilation photo-peak in the energy range $E=511 \pm 0.93\,\mathrm{keV}$  to the total peak area $\mathrm{A_{total}}$, see Fig.~\ref{fig:Definition_SW}, 
\begin{equation}
	S = \frac{A_1}{A_{total}}=\frac{\int\displaylimits^{m_ec^2+\delta E}_{m_ec^2-\delta E} N(E)dE}{\int\displaylimits^{\infty}_{-\infty} N(E)dE}
\end{equation}
where $N$ represents the number of counts in the spectrum at the energy $E$. Positrons annihilated by free electrons, which on average have a low momentum contribute to the central region of the annihilation photo-peak. An increase of vacancies will lead to a higher fraction of positrons getting trapped and annihilated by free electron in these vacancies and therefore cause an increase of the S parameter. 
\newline
A measurement of the Doppler shift of both annihilation gamma rays in coincidence enables to suppress the background below the annihilation photo-peak by three orders of magnitude. Coincidence Doppler broadening (CDB) spectroscopy~\cite{Lynn1977}, thereby, discloses the high momentum part of the momentum distribution of annihilating electron-positron pairs corresponding to contributions of core electrons. This is of interest for this study, since core electrons localized in inner shells are only weakly influenced by crystal bonding and retain their atomic character, so that their momentum distribution can be used for identification of chemical elements surrounding positron annihilation site.~\cite{Brusa2002} 

\begin{figure}[!htbp]
	\centering
		\includegraphics[width=1.00\columnwidth]{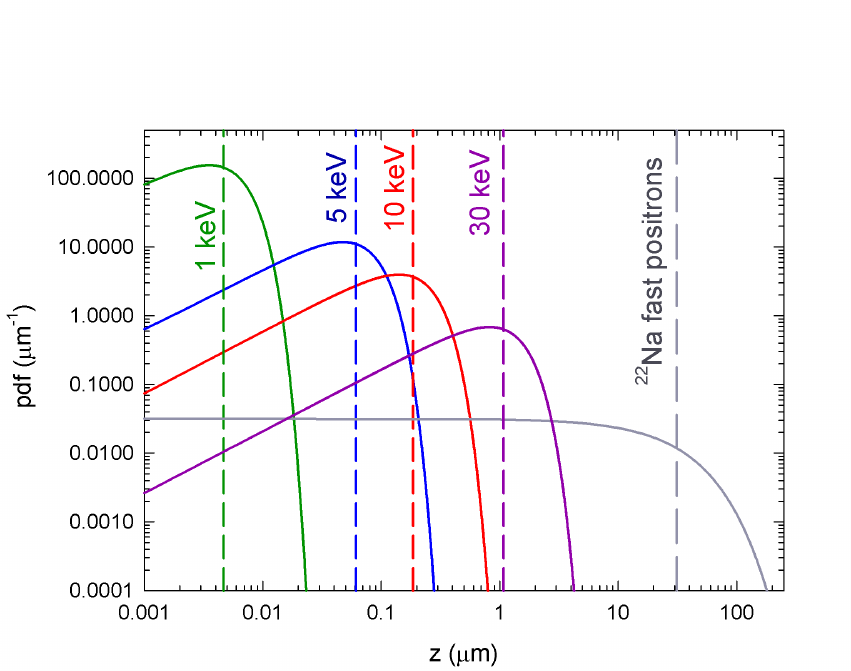}
	\caption{Implantation profile into Nb for monoenergetic positrons with energies 1, 5, 10  and 30 keV and for fast positrons with continuous energy spectrum emitted by ${}^{22}$Na radioisotope. The mean positron penetration depths are indicated by vertical dashed lines.}
	\label{fig:implanatation-profiles}
\end{figure}

Positron annihilation studies can be performed either using fast positrons with continuous energy spectrum produced by a $\beta_{+}$ radioisotope or using a variable energy slow positron beam with moderated positrons characterized by a narrow energy spectrum. 
Conventional positron annihilation spectroscopy using fast positrons provides information from the bulk of the sample while variable energy positron annihilation spectroscopy (VEPAS)~\cite{Schultz1988} using slow positron beam enables depth resolved studies of sub-surface region. Figure~\ref{fig:implanatation-profiles} shows a comparison of implantation profiles into Nb for monoenergetic positrons (used in VEPAS) with energies 1,5,10 and 30 keV and for fast positrons emitted by ${}^{22}$Na radioisotope.    

Measurements in the presented work were taken in a collaborative effort using four different setups, which were used to cross-check individual measurements but also complemented each other. Depth resolved VEPAS studies of sub-surface region were carried out at the Helmholtz Zentrum Dresden-Rossendorf on a continuous slow positron beam 'SPONSOR'~\cite{Anwand2012} and 'AIDA'~\cite{Liedke2015} employing DB spectroscopy (DB-VEPAS) as well as on a pulsed positron beam  MePS~\cite{Wagner2018, Wagner2017a} of the ELBE facility employing positron lifetime measurement (VEPALS). The spot size diameter of the positron beams were in the order of $3-5\,\mathrm{mm}$ and the energy of incident positrons was variable in the range 0.03-36 keV for DB-VEPAS (it corresponds to the mean positron penetration depth into Nb from 0.02 to 1380 nm) and 1-10 keV for VEPALS (it corresponds to the mean positron penetration depth into Nb from 4.7 to 394 nm). All VEPAS  measurements were done {\it in-situ}, i.e. samples were baked in the sample chamber of slow positron beam.
Bulk positron annihilation studies were performed at the Charles University, Prague using a high resolution digital PALS spectrometer~\cite{Becvar2005} and a digital CDB setup~\cite{Cizek2010, Cizek2012}. A ${}^{22}$Na radioisotope with an activity of 1~MBq deposited on a 2~$\mu$m Mylar foil was used as a positron source. The bulk PALS and CDB measurements were performed at room temperature and annealing treatments were done in a separate annealing chamber.  
The contribution of positrons annihilated in the positron source spot and in the covering Mylar foils was determined using a well annealed Nb reference sample. The source contribution consisted of two components: shorter component with lifetime of 368 ps and intensity 7\% and a weak long-lived component with lifetime of 1.5 ns and intensity 1\%. The shorter component comes from positrons annihilated as particles in the source spot and the Mylar foil while the long-lived component is a contribution of pick-off annihilation of ortho-Positronium formed in the Mylar foil. The source contribution was always subtracted from positron lifetime spectra of samples studied. 
\\
Bulk positron lifetime annihilation studies were carried out on a high resolution digital positron lifetime spectrometer described in \cite{Becvar2005}. The spectrometer is equipped with two scintillation detectors consisting of ${\rm BaF}_2$ scintillators coupled with fast photomultipliers Hamamatsu H3378. Detector pulses were sampled by a couple of 8-bit digitizers Acqiris DC 211 with sampling frequency of 4 GHz. Analysis of sampled waveforms and construction of positron lifetime spectrum was performed off-line using dedicated software. The spectrometer exhibits time resolution of 145 ps (full width at half maximum of the resolution function). At least $10^7$ positron annihilation events were acquired in each positron lifetime spectrum.    
\\
CDB investigations were performed on a digital spectrometer \cite{Cizek2010, Cizek2012} equipped with two HPGe detectors Canberra GC3018, GC3519. Signals from HPGe detectors were digitized using a two-channel 12-bit digitizer Acqiris DC440 which enables sampling of detector signals with sampling frequency up to 420 MHz. The energy resolution of the CDB spectrometer is 0.9 keV at the positron annihilation line. 

\subsection{\label{sec:level2.4} Software}
The data obtained at MePS was analyzed using the PALSfit3 software \cite{PALSfit3} and the plots were generated using OriginLab \cite{OriginLab}. The data obtained at the positron annihilation group at the Charles University was analyzed using dedicated software developed in the Positron annihilation group at the Charles University. The software uses integral constant fraction technique \cite{Becvar2007} and is described in \cite{Becvar2005}. Decomposition of positron lifetime spectra into individual components was performed using a maximum likelihood-based code described in \cite{Prochazka1997}. These plots and the nanoindent plots were generated using Sigma Plot \cite{SigmaPlot}. 

\section{\label{sec:level3}Results}
\subsection{Sub-surface positron annihilation studies}
\label{VEPAS-section}
\begin{figure}[!htbp]
	\centering
		\includegraphics[width=1.00\columnwidth]{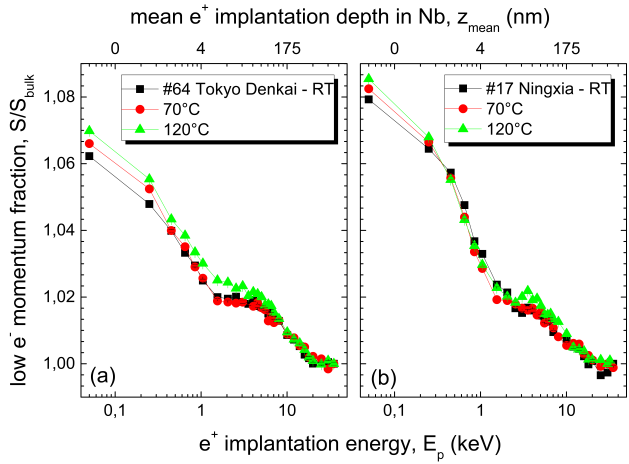}
	\caption{S parameter normalized to the bulk value $S_{\rm bulk}$ as a function of positron implantation energy $E_{\rm p}$ and the mean positron penetration depth $z_{\rm mean}$ measured by means of DB-VEPAS and annealed {\it in-situ} at different temperatures on samples: (a) 64 - Tokyo Denkai, and (b) 17 - Ningxia.}
	\label{fig:S_E_64}
\end{figure}

The results of DB-VEPAS measurements of the sample 64 (Tokyo Denkai) are shown in Fig.~\ref{fig:S_E_64}a The sample 17 (Ningxia) shown in Fig.~\ref{fig:S_E_64}b  exhibits the same trend. The S parameters are normalized to the bulk value $S_{\rm bulk}$ (measured for a reference sample of well annealed iron at the positron implantation energy of 35 keV corresponding to the mean penetration depth of $1.5\;\upmu$m) and plotted as a function of the positron implantation energy $E_{\rm p}$ and the mean positron penetration depth $z_{\rm mean}$ for three distinct temperatures: RT, $70^\mathrm{o}\,\mathrm{C}$ and $120^\mathrm{o}\,\mathrm{C}$. The S-parameter decreases monotonically as a function of $E_{\rm p}$ exhibiting, however, three distinct regions across sample thickness: (i) surface layer for $E_{\rm p}<$\,1\,keV (5~nm) consisting likely of niobium oxides, (ii) sub-surface region for 1\,keV (5~nm)\,$<E_{\rm p}<$\,10\,keV (190~nm) representing expected vacancy-hydrogen complexes, and (iii) Nb bulk. 

For all samples, the observable changes happen in a sub-surface region with thickness of about $\approx$ 190\,nm. This is sufficient to influence the rf performance, since the London penetration depth $\uplambda_L$ is 40-50\,nm for clean niobium at 2\,K. All samples show an increase of the S parameter in the sub-surface region during the baking for 4\,h up at  $120^\mathrm{o}\,\mathrm{C}$, which is an indication of increased open volume and/or concentration of defects. 

\begin{figure}[!htbp]
	\centering
		\includegraphics[width=1.00\columnwidth]{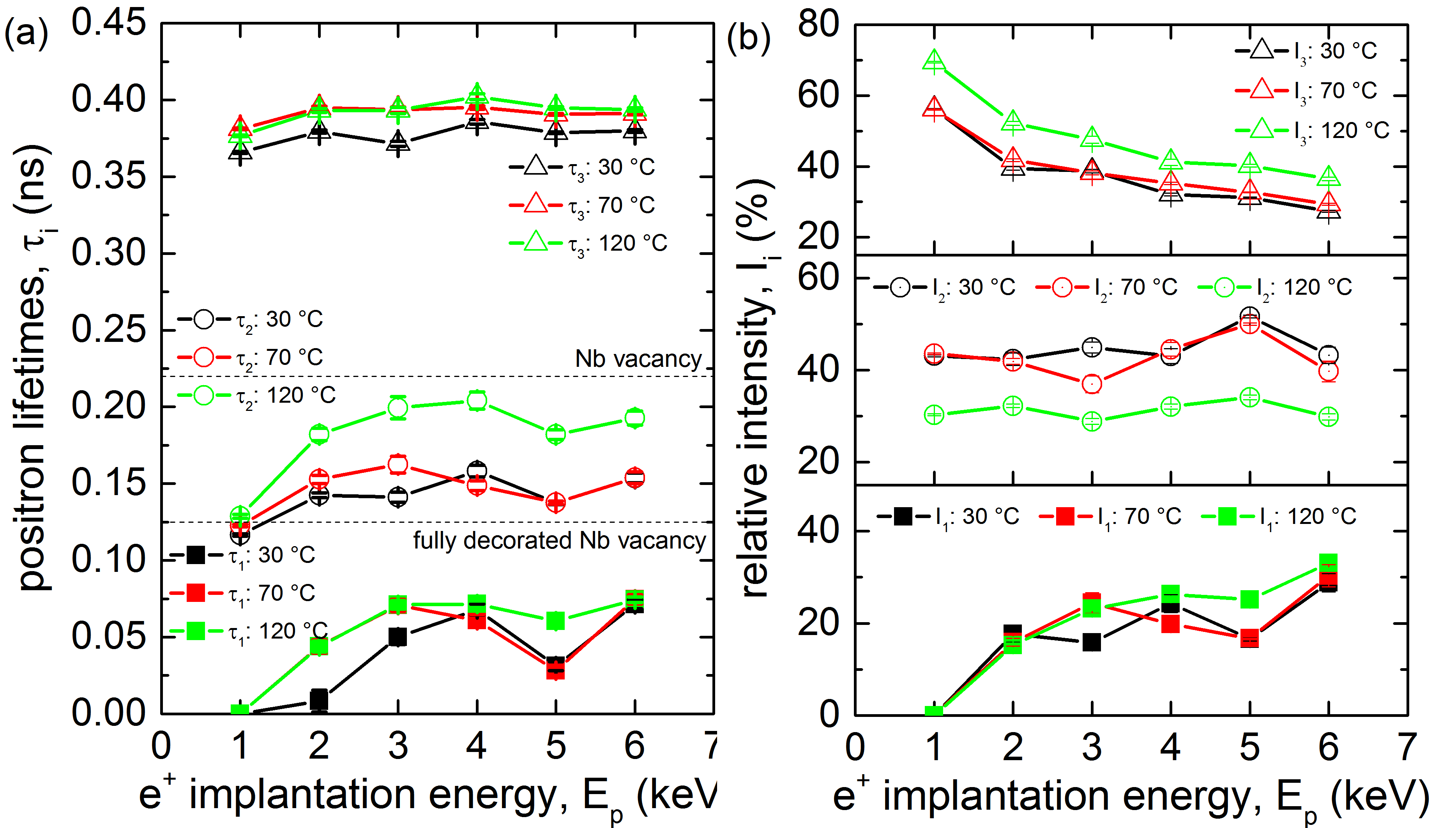}
	\caption{Positron lifetime components $\tau_1$, $\tau_2$ and $\tau_3$ (a) as well as relative intensities $I_1$-$I_3$ of the lifetime components (b) as a function of positron implantation energy for the sample 78 measured {\it in-situ} before and after baking at $70^\mathrm{o}\,\mathrm{C}$ and $120^\mathrm{o}\,\mathrm{C}$ for 4~ h.}
	\label{fig:ELBE_78}
\end{figure}

Results of depth resolved VEPALS measurement, i.e. the dependence of positron lifetimes $\tau_1$, $\tau_2$ and $\tau_3$ of exponential components resolved in positron lifetime spectra on the energy of incident positrons, for the sample 78 (Tokyo Denkai) are shown in ~Fig. \ref{fig:ELBE_78}a.
Corresponding intensities are plotted in Fig.~\ref{fig:ELBE_78}b. Note that relative intensities are normalized so that $I_1 + I_2 + I_3 = 100$\,\%. 
The Positron lifetime spectra near the surface consisted of three components: (i) a short-lived component with lifetime $\tau_1$ and intensity $I_1$ representing a contribution of free positrons delocalized in the lattice (not trapped at defects). 
(ii) a medium-lived component with lifetime $\tau_2$ and intensity $I_2$ which can be attributed with positrons trapped at v+nH complexes.
(iii) a long-lived component with lifetime $\tau_3$ and intensity $I_3$ which can be attributed to positrons trapped positrons trapped at larger defects, e.g., vacancy clusters, and positrons annihilated in the surface state.

The lifetime $\tau_1$ of the free positron component is shorter than the bulk positron lifetime for niobium of 125\,ps \cite{Cizek2004} due to positron trapping at defects regardless of the baking, see Fig.~\ref{fig:ELBE_78}a.  These findings indicate that the samples contain open volume defects.
In view of bulk positron lifetime studies described in the section~\ref{bulk-PAS} it is likely that samples in the sub-surface region contain mono-vacancies, associated with hydrogen and positrons trapped at these defects contribute to the component $\tau_2$. Comparing the calculations shown in Fig. \ref{fig:lifetime_vnh} with the observed lifetimes, vacancies in the sub-surface region are occupied by 4 or more hydrogen atoms at once.  
\\
The lifetime $\tau_3$ is significantly higher than lifetime of positrons trapped in mono-vacancies in Nb (222 ps \cite{Cizek2004}). The lifetime $\tau_3$ corresponds to  positrons trapped at larger defects, namely vacancy clusters consisting of several vacancies.
The component $\tau_3$ contains also a contribution from positrons annihilated in the surface state, i.e. positrons confined in a potential well on the surface determined by the outermost atomic layer and the surface image potential of the positron~\cite{Hugenschmidt2016,Chu1981}
With increasing energy positrons penetrate deeper and deeper into the sample and the fraction of positrons diffusing back to surface decreases. 
The intensity $I_3$ decreases with increasing positron energy since the number of positrons trapped in the surface states decreases and therefore more positrons are annihilated in the free states or v+nH complexes with $\mathrm{n}\geq4$.

Baking at $70^\mathrm{o}\,\mathrm{C}$ changed the defect distribution and composition across the sample thickness. The lifetime components $\tau_1$ and $\tau_2$ below 3\,keV increased while the lifetime component $\tau_3$ increased all over the penetration depth, see  Fig.~\ref{fig:ELBE_78}a. 
This indicates an increase of the vacancy clusters free volume in the sub-surface region while their concentration remains approximately unchanged. Such a scenario is possible, if vacancy clusters in the original sample were decorated by hydrogen and baking at $70^\mathrm{o}\,\mathrm{C}$ caused release of hydrogen from the clusters. It is known that hydrogen trapped at open volume defects reduces lifetime of trapped positrons \cite{Cizek2004}. Hence, release of hydrogen from the clusters leads to increase of positron lifetime. Baking at $120^\mathrm{o}\,\mathrm{C}$, on the other hand, exhibits increased positron annihilation at vacancy clusters and surface states since $I_3$ raises to about 70\,\%. It is only possible, if the overall number of mono-vacancies associated with hydrogen atoms drops substantially due to annealing. The size of vacancy clusters does not change much ($\tau_3$) but because of lower concentration of vacancy-like defects positrons can diffuse longer reaching the surface. The lifetime $\tau_1$ remains mostly unaffected in the near surface region below 4\,keV after baking at $120^\mathrm{o}\,\mathrm{C}$ for 4\,h. On the other hand, $\tau_2$ increases substantially while $I_2$ decreases. The increase of $\tau_2$ strongly suggests that v+nH complexes dissociated, i.e. H atoms separated from mono-vacancies. The lifetime change suggests that while 4 or more hydrogen atoms were associated with vacancies before baking, only 1 to 2 atoms per vacancy remain on average attached to vacancies after baking at $120^\mathrm{o}\,\mathrm{C}$ for 4\,h.  

\begin{figure}[!htbp]
	\centering
		\includegraphics[width=1.00\columnwidth]{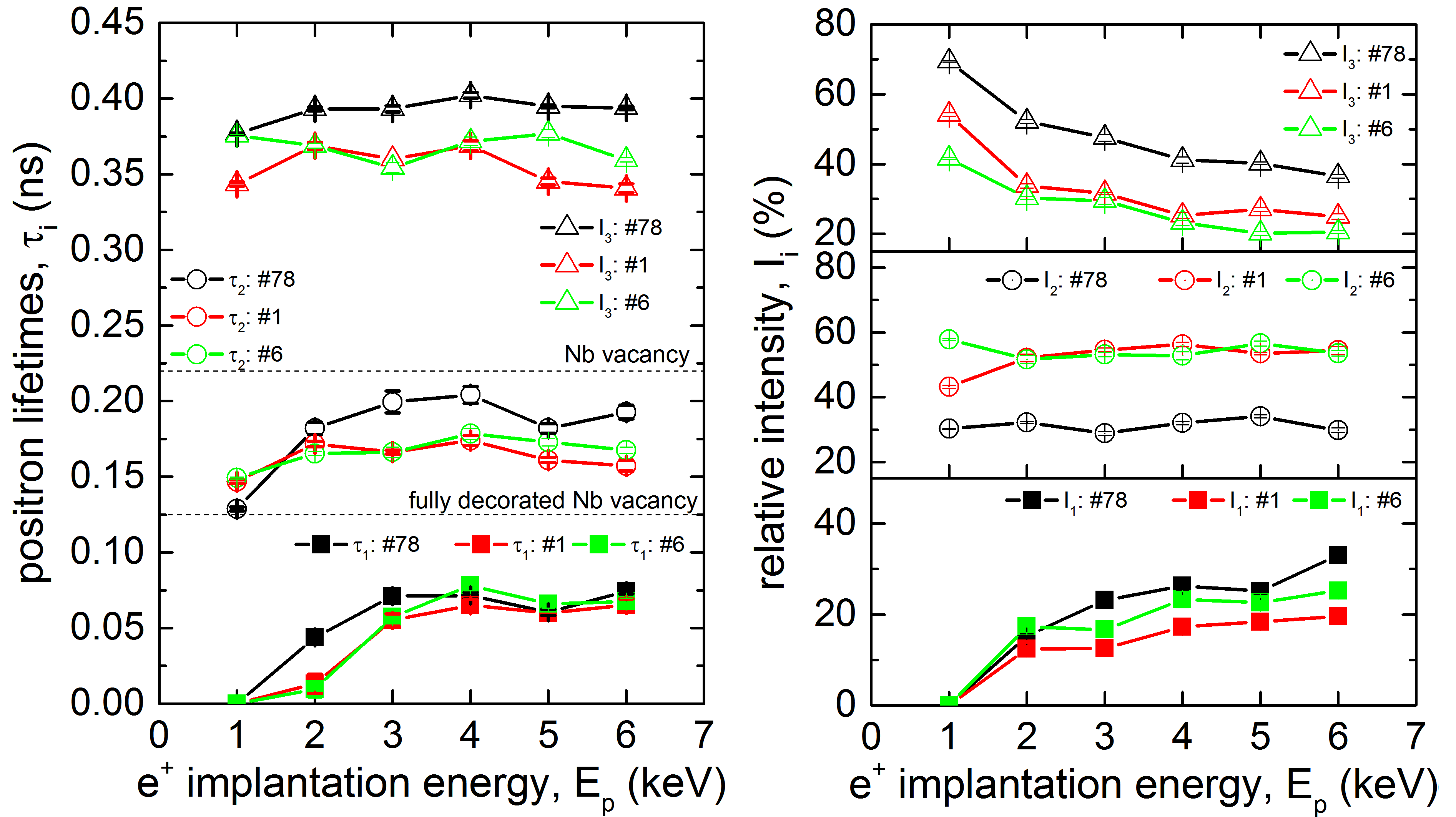}
	\caption{Lifetimes $\tau_1$, $\tau_2$ and $\tau_3$ (left) and the intensities $I_1$-$I_3$ of positrons trapped at defects (right) plotted as a function of positron energy. The data were measured at room temperature for a set of cavity cut outs and the data from sample 78 is shown for comparison.}
	\label{fig:ELBE_CutOuts_three}
\end{figure}

\begin{figure}[!htbp]
	\centering
		\includegraphics[width=1.00\columnwidth]{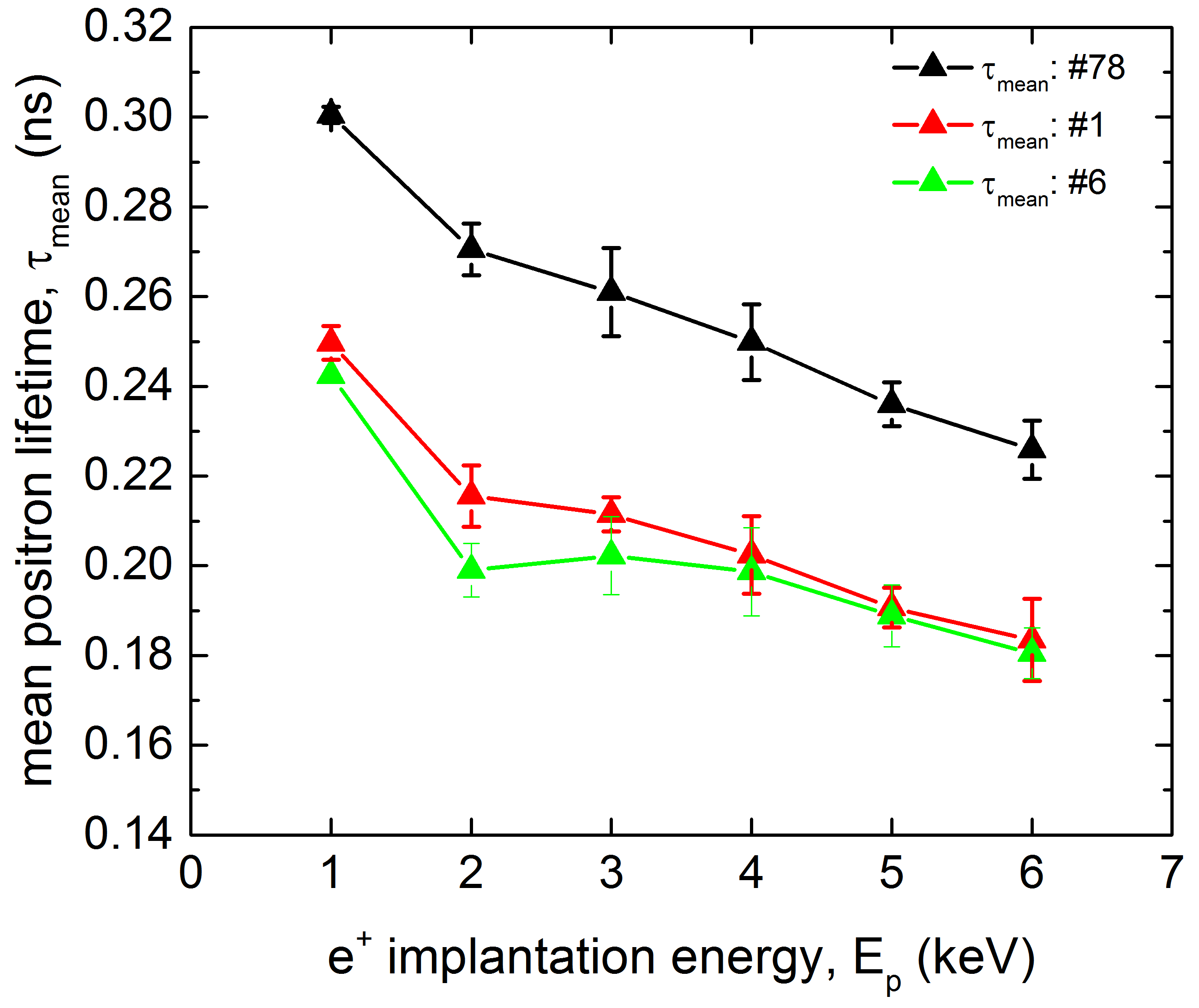}
	\caption{Mean lifetimes $\tau_{mean}$ of positrons trapped at defects plotted as a function of the positron energy for three samples: cavity-cut outs \#1 and \#6  and sample 78.}
	\label{fig:ELBE_CutOuts_Mean}
\end{figure}

In order to crosscheck the sample measurements with real cavity surfaces, measurements of samples cut-outs (cuts No. \#1 and \#6) from a tested cavity were done at MePS and the results are shown in Fig. \ref{fig:ELBE_CutOuts_three}, supporting the results obtained for the sample 78. The cavity underwent the same preparation as the samples described in section \ref{sec:sample_prep} but the thickness of the layer removed by EP was higher by $30\, \upmu \mathrm{m}$. In addition, the final bake for 48\,h at $120^\mathrm{o}\,\mathrm{C}$ in a pressure of $\leq1\times10^{-5}\,\mathrm{mbar}$ was applied. During the chemical treatment, a surface layer of $180\, \upmu \mathrm{m}$ was removed, which is $30\, \upmu \mathrm{m}$ more than for the samples.    
The overall trends for both types of samples agree well. The observed difference in the absolute values most likely comes from the higher concentration of hydrogen absorbed in the lattice for the cut out samples, gathered over the different treatments and bakings. The lower density of large vacancy-clusters, described by $I_3$, in the cut out samples is a consequence of the larger amount of removed material by $30\, \upmu \mathrm{m}$ from the surface and hence, less remaining damage layer from the mechanical formation. The lifetime component $\tau_2$ of the cut outs of around 150\,ps shows that the vacancies are still associated with 4 or more hydrogen atoms. This difference can be explained by the fact that the samples cut outs were baked in vacuum two orders of magnitude worse than that used in MePs beam for measurement of samples baked in-situ. Hence, the hydrogen concentration in the real cut outs might be higher then in model samples baked in the MePs apparatus.
\newline
Another question which was investigated with MePS on the cut outs was the origin of the cavity limitation during operation. The cavity underwent a test at 2\,K to measure the surface resistance as a function of the applied accelerating field before cutting. Allen-Bradley Resistors were glued to the outside of the cavity to measure the local heating over the inner cavity surface while applying the field. The spatial resolution of the temperature map was $9\times13\,\mathrm{mm}$. A \text{Fein Multimaster} was used to cut the square-shaped samples, with a side length of $\approx\mkern-8mu 1.5\,\mathrm{cm}$. Sample 1 was cut from the limiting region in the cavity test, which means that the cavity underwent the phase transition from superconducting to normal state, so called \textit{quench}, at only 60\% of the maximal critical field. Optical inspections showed no defect on the surface of sample 1 but it was found that 14\,mG of the external field was trapped in this quench region, which is 10-15 times higher than the average value for a cavity \cite{Wenskat2019b}. Sample 6 was cut from a region which showed average temperature and rf behavior during testing \cite{Wenskat2019b}. The mean lifetimes as a function of the positron energy of those samples are shown in Fig. \ref{fig:ELBE_CutOuts_Mean}. The overall behavior of the mean positron lifetime for cut outs 1, 6 is similar to that for sample 78 but the mean lifetimes for the cutouts are shifted to lower values due to lower density of vacancy clusters
An interesting observation is the difference between the sample 1 and samples 6 in the positron implantation energy range of 1-3\,keV in the sub-surface region 14 - 40~nm. The mean lifetime $\tau_{mean}$ of sample 1 exhibits here an excess compared to sample 6 representing increased open volume of vacancy cluster and near-surface defects. Lattice defects such as  grain boundaries, vacancies and interstitials are known to be pinning sites for flux lines when the material undergoes the phase transition and the magnetic flux is expelled \cite{Antoine2019a}. Hence, it is likely that the origin of the cavity quench originates from vacancy clusters found in the sample 1. These vacancy clusters act as strong pinning sites and trap  significant amount of flux which then causes an additional surface resistance, leading to high thermal losses and limiting the rf performance. The results for cavity cut out further strengthens the credibility of the applied methods in order to understand observed rf performance of real cavities.  

\subsection{\label{bulk-PAS}Bulk positron annihilation investigations}
\begin{figure}[!htbp]
	\centering
		\includegraphics[width=1.00\columnwidth]{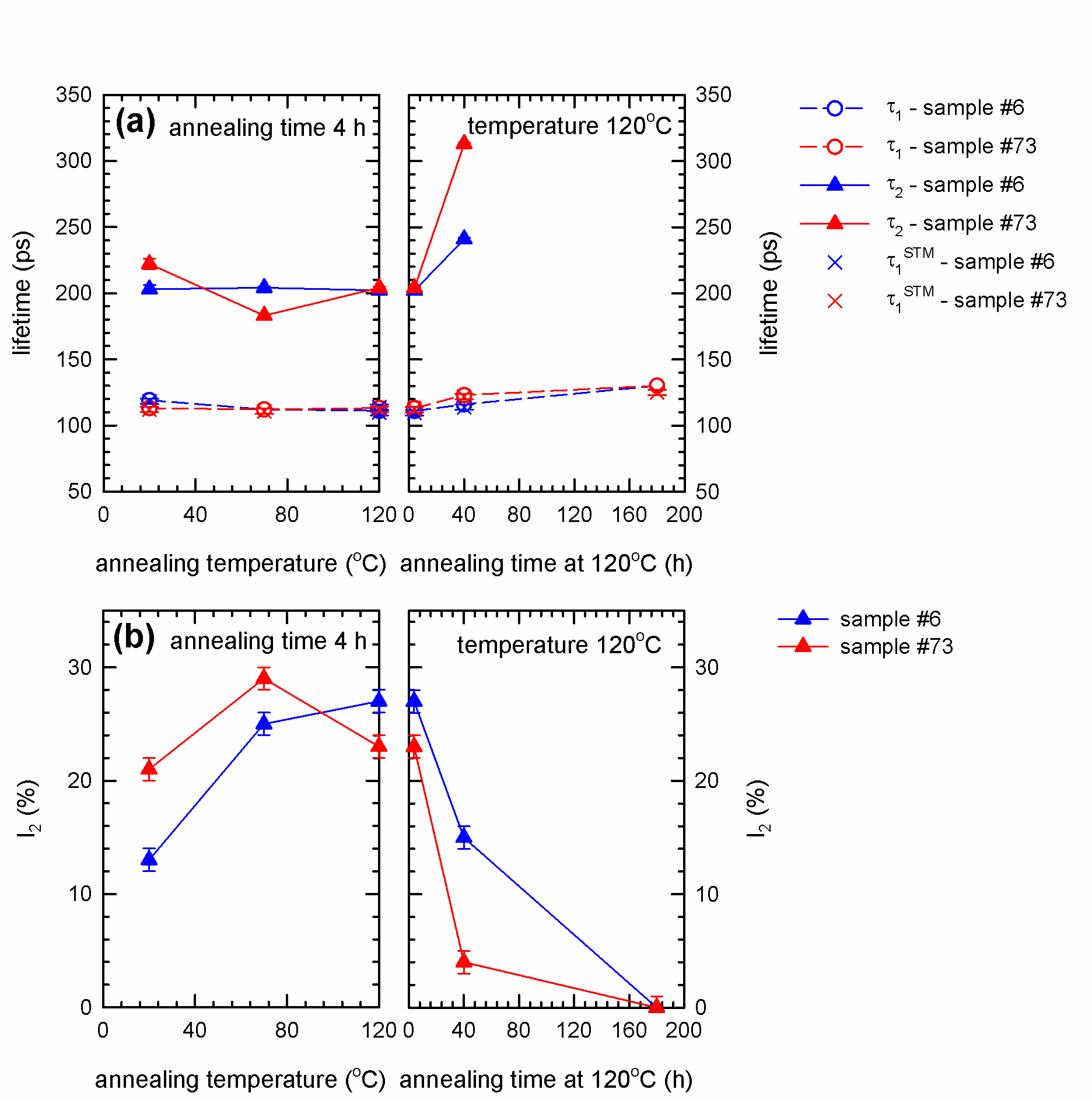}
	\caption{Results of bulk PALS measurements of samples annealed {\it ex-situ} in a vacuum furnace: temperature dependence of (a) lifetimes of exponential components resolved in LT spectra. The free positron lifetime $\tau_1^{STM}$ calculated within the two state trapping model is shown in the figure as well; (b) relative intensity $I_2$ of positrons trapped at defects.}
	\label{fig:Lifetime}
\end{figure}

\begin{figure}[!htbp]
	\centering
		\includegraphics[width=1.00\columnwidth]{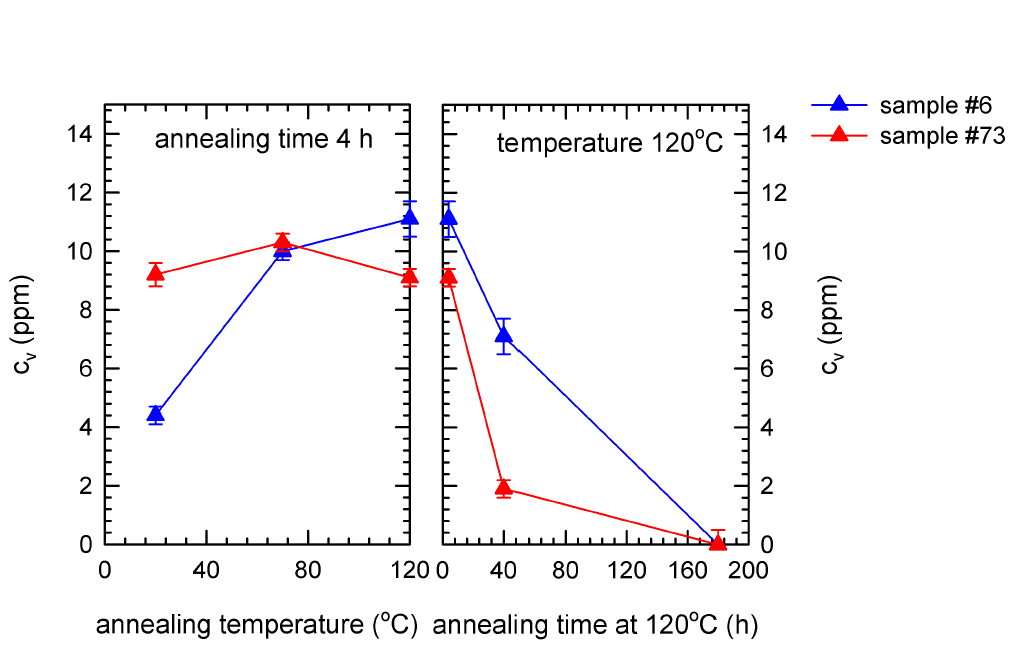}
	\caption{Vacancy concentrations in bulk calculated from PALS data in \ref{fig:Lifetime} using two state positron trapping model vs. annealing temperature.}
	\label{fig:VacConc}
\end{figure}

The results of the bulk PALS measurements for samples 6 (Ningxia) and 73 (Tokyo Denkai) are shown in fig. \ref{fig:Lifetime} and \ref{fig:VacConc}.  
\newline
The spectra for bulk region can be well described by two exponential components. The shorter component with lifetime $\tau_1$ and intensity $I_1$ stems from free positrons while the longer component with lifetime $\tau_2$ and intensity $I_2=100\% -I_1$ represents a contribution of positrons trapped at open volume defects.  
The lifetime $\tau_2$ is around 200\,ps for the virgin sample.  This value is shorter than the lifetime of positrons trapped at bare mono-vacancies (222\,ps \cite{Cizek2004}) and corresponds well to vacancies associated with single hydrogen atom (v+H), see Fig.~\ref{fig:lifetime_vnh}.  Hence bulk PALS measurements revealed that Nb samples from both suppliers contain in bulk vacancies associated with hydrogen. 
During annealing up to $120^\mathrm{o}\,\mathrm{C}$ the lifetime $\tau_2$ remains approximately constant but the intensity $I_2$ increases indicating that the concentration of v+H increases with annealing temperature up to $120^\mathrm{o}\,\mathrm{C}$. 

The samples were subsequently annealed at $120^\mathrm{o}\,\mathrm{C}$ for various time periods (see right panel of Fig. \ref{fig:Lifetime}). The  prolonged annealing at $120^\mathrm{o}\,\mathrm{C}$ leads to an increase of the lifetime $\tau_2$ up to $\sim 300$~ps indicating that vacancies agglomerate into clusters. This is accompanied by a decrease of the intensity $I_2$ which testifies that the concentration of vacancies decreases with a longer annealing period at $120^\mathrm{o}\,\mathrm{C}$. Hence, PALS results show that not only higher temperatures but also longer baking time has important effect on the defect evolution. 

The free positron lifetime 
\begin{equation}
    \tau_1^{STM} = \frac{I_1}{\tau_B^{-1}-I_2\tau_2^{-1}}
    \label{tau_1-STM}
\end{equation}
calculated within the two-state simple trapping model~\cite{West1973} is plotted in Fig.~\ref{fig:Lifetime} as well and agrees well with the lifetime $\tau_1$ measured in experiment. 
It testifies that the assumptions of two state simple trapping model are fulfilled in the present case and it can be used for determination of defect concentration.
The concentration of vacancies (or vacancy clusters) calculated using the two state simple trapping model by Eq.~(\ref{cv}) is plotted in Figure \ref{fig:VacConc}.  The concentration of vacancies increases for short baking periods 4~h at temperatures below $120^\mathrm{o}\,\mathrm{C}$ but a long-term baking at $120^\mathrm{o}\,\mathrm{C}$ leads to a decrease of vacancy concentration. Although there is certain difference between samples 6 and 73 (initial concentrations in the sample 6 is lower) the overall behavior of both samples is quite similar. 

\begin{figure}[!htbp]
	\centering
		\includegraphics[width=1.00\columnwidth]{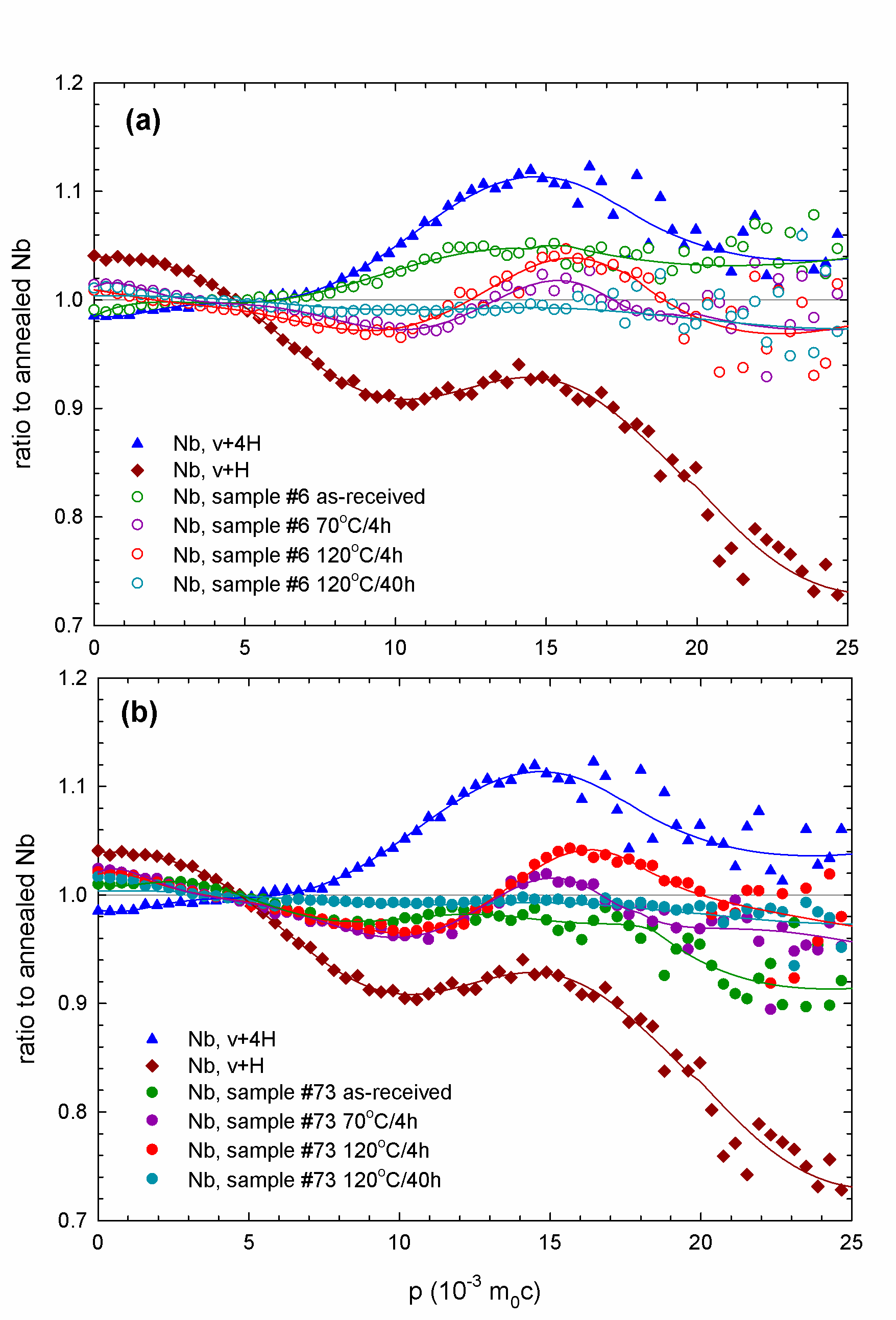}
	\caption{CDB ratio curves related to a well annealed ($1000^\mathrm{o}\,\mathrm{C}$ in UHV) Nb reference  for (a) sample 6 (Ningxia), (b) sample 73 (Tokyo Denkai). Two reference curves for Nb with v+H and v+4H complexes are shown for comparison. }
	\label{fig:CBD}
\end{figure}

To better analyze the interaction of vacancies with hydrogen in the samples, bulk CDB measurements were done. The CDB ratio curves for the samples 6 (Ningxia) and 73 (Tokyo Denkai) are  shown in Figs. \ref{fig:CBD}a and \ref{fig:CBD}b, respectively. The CDB ratio curves are related to a well annealed ($1000^\mathrm{o}\,\mathrm{C}$ in UHV) Nb reference exhibiting a defect-free lattice. The CDB curves measured in Nb samples are superpositions of the contribution of positrons trapped at v+nH complexes and the free positron contribution which is a horizontal line in unity. Hence, with decreasing concentration of v+nH complexes the CDB curves becomes closer to unity.  Two CDB curves of reference samples containing vacancy-hydrogen (v+H) complexes and vacancies associated with 4 hydrogen atoms (v+4H) are shown in the figure as well. These reference samples were prepared by electron irradiation in order to introduce vacancies and subsequent hydrogenation \cite{Cizek2009}. The ratio curves of reference samples containing vacancy-hydrogen complexes have a characteristic peak at $p = 14 \times 10^{-3} \mathrm{m_0 c}$ which represents a contribution of positrons annihilated by 1s hydrogen electrons \cite{Cizek2009}. The height of the peak at $14 \times 10^{-3} m_0c$ increases with increasing number of hydrogen atoms attached to vacancy \cite{Cizek2009}. 
While for the as-received sample, some hydrogen is already associated to vacancies v+nH complexes form at $70^\mathrm{o}\,\mathrm{C}$ and are stable up to $120^\mathrm{o}\,\mathrm{C}$ with short annealing time of 4 h. After long-term annealing at $120^\mathrm{o}\,\mathrm{C}$ no v+nH complexes are observed in the bulk anymore.
\begin{figure}[!htbp]
	\centering
		\includegraphics[width=1.00\columnwidth]{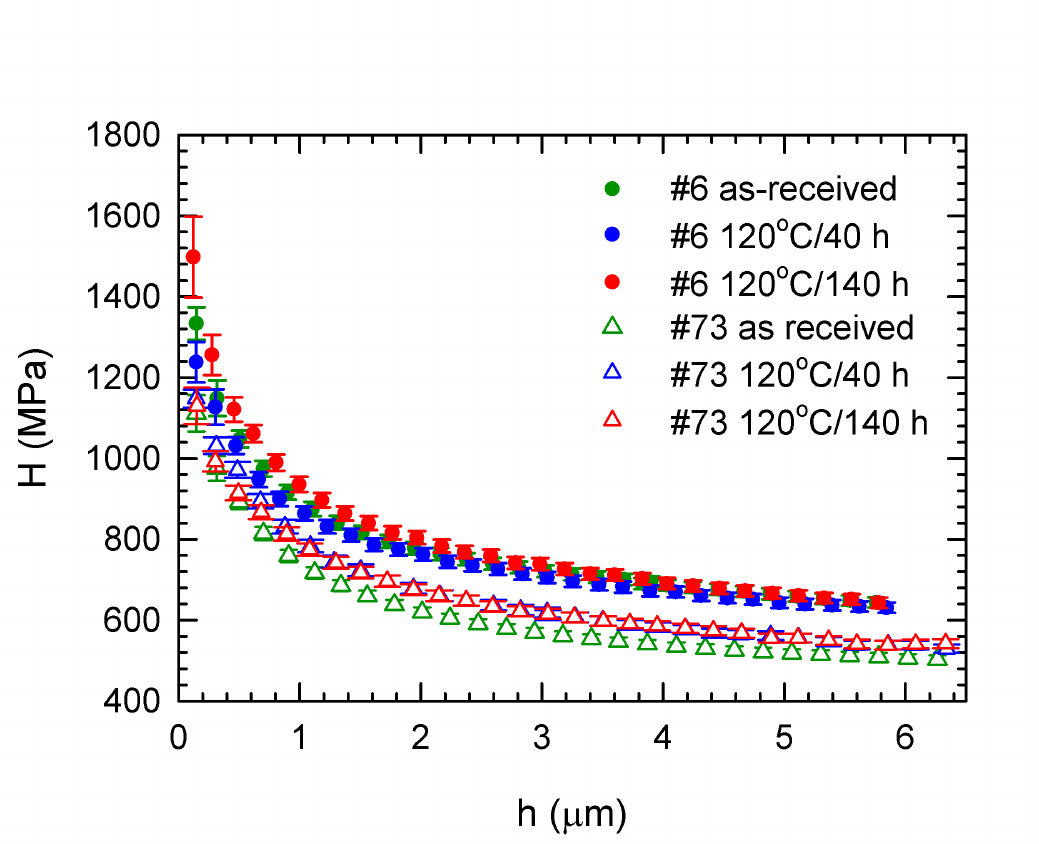}
	\caption{Dependence of hardness on the maximum indentation depth. Although sample 6 shows an overall higher hardness than sample 73, the shape of the curves is similar, pointing towards a similar density of obstacles for motion of dislocations.}
	\label{fig:Indent_before}
\end{figure}
\begin{figure}[!htbp]
	\centering
		\includegraphics[width=1.00\columnwidth]{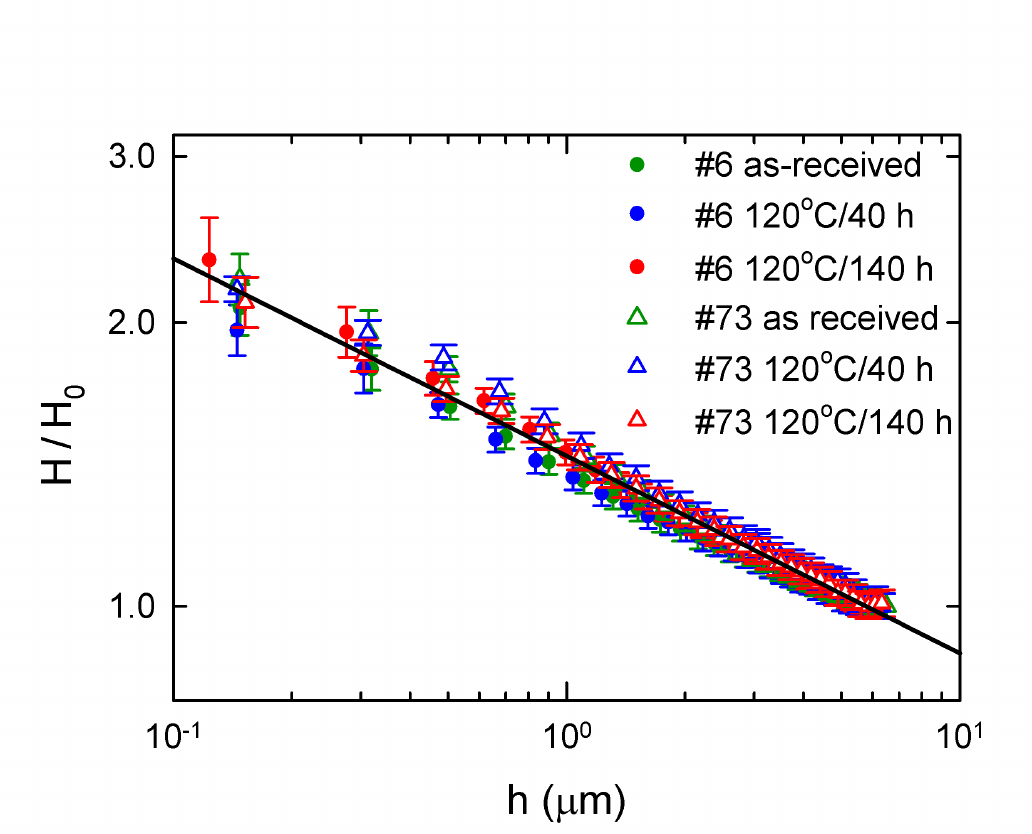}
	\caption{Log-log plot of hardness versus indentation depth. The hardness $H$ measured at various depths from 150~nm up to 5~$\mu$m were divided by the bulk value of hardness $H_0$ measured using the loading force of 500~nN for which the indentation depth exceeded 5\,$\mu$m.}
	\label{fig:log(H)-Log(h)}
\end{figure}
\begin{figure}[!htbp]
	\centering
		\includegraphics[width=1.00\columnwidth]{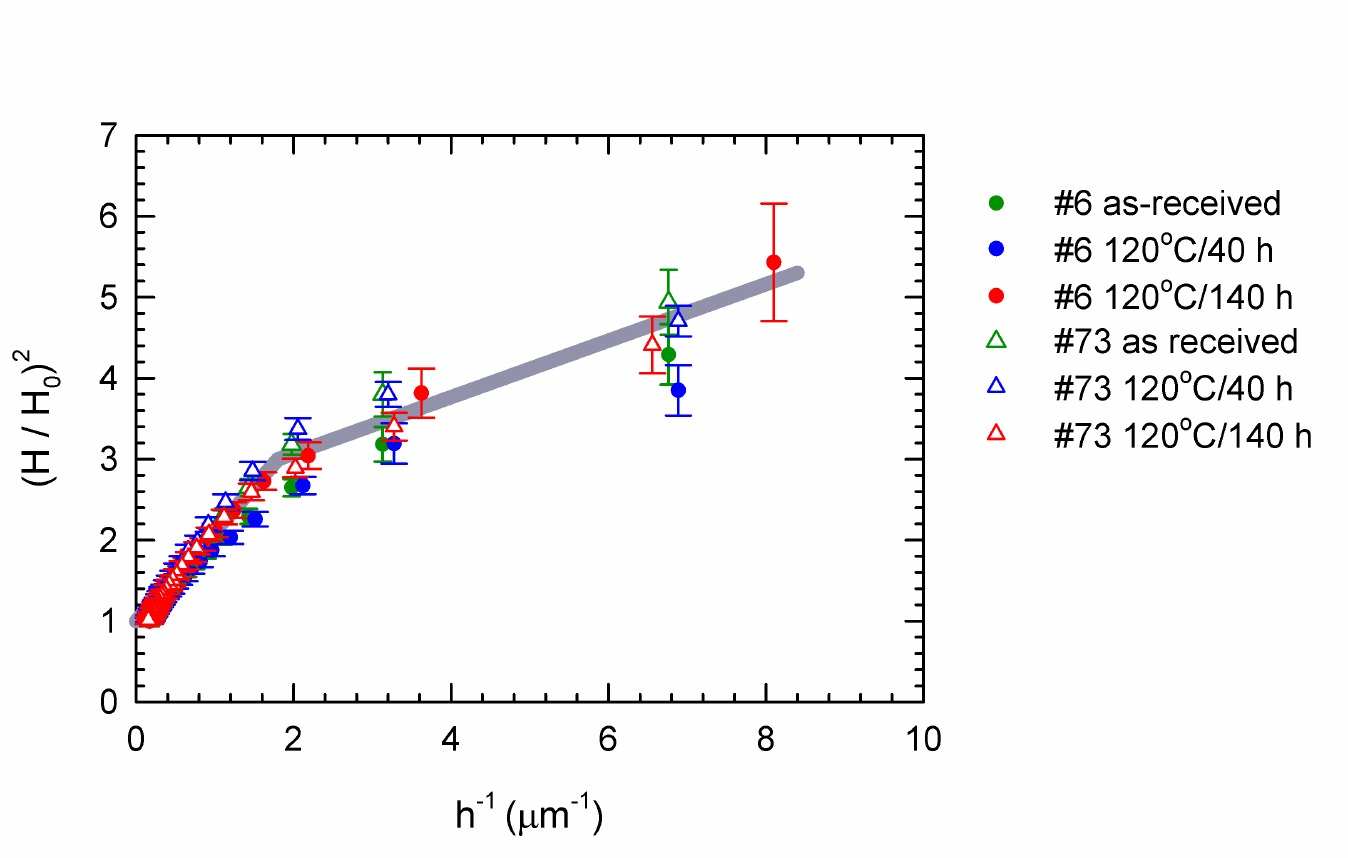}
	\caption{$(H/H_0)^2$ plotted as a function of $1/h$. Solid line shows results of fitting using the Kasada's model~\cite{Kasada2011} }
	\label{fig:Nix-Gao-test}
\end{figure}
\subsection{Nanoindentation}
Fig.~\ref{fig:Indent_before} shows the results of instrumented nanoindentation. The sample 6 exhibits higher hardness values than the sample 73. This is in accordance with higher concentration of interstitial impurities (C,N,O,H) in this sample. The macroscopic HV values estimated from nanoindetation (H) curves correspond to HV $\approx$ 47 and 60 for sample 73 (Tokyo Denkai) and 6 (Ningxia), respectively. The HV value determined for the sample 73 fits well into the range of 42.5-50.6 provided by the producer (Tokyo Denkai), while the sample 6 exhibits a HV value remarkably higher than the value range 37-46 specified by the producer (Ningxia).  
The hardness $H$ increases with decreasing indentation depth $h$.
This so called indentation size effect (ISE)~\cite{Nye53,Milman11} is known to occur in single crystals or coarse grained polycrystals typically for an indentation depths bellow 10~$\upmu$m. Various mechanisms were proposed to explain this phenomenon. The ISE was related to the change in the contribution of the elastic and plastic deformation in the indentation region~\cite{Dub2010}, the strain gradient~\cite{Nix1998}, the effect of surface~\cite{Gerberich02}, etc.    
Experimental data for various materials~\cite{Manika06, Milman11} including single crystals of metals and semiconductors as well as coarse grained metallic polycrystals generally show a power law relation between hardness $H$ and the indentation depth $h$      
\begin{equation}
H = \frac{C}{h^{m}}, 
\end{equation}
where $C$ is a constant and $m$ is the power-law exponent or ISE index~\cite{Manika06}. 
Fig.~\ref{fig:log(H)-Log(h)} shows a plot of ${\rm log} (H/H_0)$ versus ${\rm log}\, h$ for samples 6 (Ningxia) and 73 (Tokyo Denkai). Points for all samples fall on a common straight line with slope $m = 0.22\pm 0.01$. This value of the ISE index fits well into the observed range of 0.15 -- 0.28  determined for various cubic metals~\cite{Manika06}

Indentation results in a plastic flow of the material which can be considered as a shear along semicircular slip surfaces with a center in the axis of the indenter. This results in formation of geometrically necessary dislocations (GNDs) which cause additional hardening~\cite{deGuzman93}. Nix and Gao~\cite{Nix1998} developed a simple model assuming that indentation is accommodated exclusively by circular loops of GNDs with a Burgers vector perpendicular to the sample. Within this model hardness is related to the indentation depth by the relation  
\begin{equation}
\frac{H}{H_0} = \sqrt{1+\frac{h^{\star}}{h}},
\label{Nix-Gao}
\end{equation}
where $H_0$ is the hardness in the absence of GNDs, i.e. hardness determined in sufficiently high depth so that density of GNDs is negligible compared to the density of dislocations statistically stored in the material. The quantity $h^{\star}$ is a characteristic length depending on the shape of indenter, shear modulus of the material and $H_0$. Hence, the characteristic length $h^{\star}$ depends on the density of statistically stored dislocations through its dependence on $H_0$. Consequently, for a given material, $h^{\star}$ is a measure of the spacing of obstacles of dislocation slip.  
Fig.~\ref{fig:Nix-Gao-test} shows $(H/H_0)^2$ plotted as a function of $1/h$, so called Nix-Gao plot. It is clear that relation~(\ref{Nix-Gao}) is not full filled in the present case since the dependence of $(H/H_0)^2$ on $1/h$  exhibits at least two slopes higher slope in high indentation depths (small $1/h$ values) and lower slope in small indentation depths (high $1/h$ values). Hence, ISE in the Nb samples studied cannot be explained solely by the Nix and Gao model. 
There are several possible explanation of lowering of the slope of the Nix-Gao plot at low indentation depths (high $1/h$ values):\\ (i) {\it sub-surface region with enhanced hardness.} The shape o the Nix-Gao plot in Fig.~\ref{fig:Nix-Gao-test} is qualitatively similar to those reported for ion irradiated metals~\cite{Kasada2011,Kasada2014,Kasada2016,Hasenhuetl2017} consisting of harder sub-surface region containing radiation-induced defects and softer bulk. VEPAS investigations described in Section~\ref{VEPAS-section}  revealed enhanced concentration of vacancies and vacancy clusters associated with hydrogen in the sub-surface region of Nb samples studied. Hence, in analogy with ion irradiated materials, one can assume that Nb samples consist of harder sub-surface region with enhanced concentration point defects and softer bulk region. This results in two slopes in the Nix-Gao plots for Nb samples in Fig.~\ref{fig:Nix-Gao-test}. The data in Fig.~\ref{fig:Nix-Gao-test} can be, indeed, well described using the model developed by Kasada et al.~\cite{Kasada2011} for ion irradiated materials, see solid lines in in Fig.~\ref{fig:Nix-Gao-test}. Analysis of data in Fig.~\ref{fig:Nix-Gao-test} within the Kasada's model~\cite{Kasada2011} revealed that the bulk equivalent hardness in the sub-surface region is $\approx 1.5$ times higher than the bulk hardness $H_0$. This harder  sub-surface region has depth of $\approx 0.5$~$\mu$m and exhibits  the characteristic length $h^{\star}\approx 160$~nm which is roughly 5 times shorter than the characteristics length $h^{\star}\approx 800$~nm in the bulk region. This indicates that the sub-surface region contains enhanced density of obstacles for motion of dislocations in accordance with enhanced concentration of v+nH complexes and vacancy clusters determined by VEPAS.\\
(ii) {\it maximum allowable GND density.} In the Nix-Gao model~\cite{Nix1998} the density of GNDs monotonically increases with decreasing indentation depth and should be very high for very small indentation depths. However, the actual GND density cannot be very large because repulsive forces between GNDs cause spread of dislocations beyond the hemisphere underneath the indentor~\cite{Huang2006}. In certain critical indentation depth $h_{nano}$ the density of GNDs reaches the maximum allowable value $\rho_{GND,max}$ and does not increase with further reduction of indentation depth. As a consequence the slope of Nix-Gao plot decreases~\cite{Huang2006,Ruiz-Moreno2018}. Assuming $\rho_{GND,max}\sim 10^{16}$~${\rm m}^{-2}$ representing typical value for metals~\cite{Huang2006}, one obtains the critical indentation depth $h_{nano}\sim 100$~nm~\cite{Huang2006}.  The effect of maximum allowable density of GNDs certainly reduces the slope of the Nix-Gao plot of Nb samples studied but it occurs likely at indentation depths significantly lower than 0.5~$\mu$m ($1/h > 2\;\mu\text{m}^{-1}$) where the change of the slope of the Nix-Gao plot in Fig.\ref{fig:Nix-Gao-test} was observed.\\
Hence, one can conclude that slope of the Nix-Gao plot of the Nb samples is probably reduced by both effects (i) and (ii). Enhanced hardness of the sub-surface region indicates that there is a hardening mechanism related to vacancies and hydrogen. 
Molecular dynamic simulations performed for Fe revealed that bare vacancies are absorbed by gliding dislocations, but vacancies associated with hydrogen are not dissociated even when meeting dislocation sinks \cite{Li2015}. It is likely that vacancies in Nb behave in similar way. Since hydrogen binding energy to vacancy in Nb is rather high (0.5 eV) \cite{Cizek2004} v+nH complexes are assumed to be stable against decomposition when meeting dislocation sinks and may represent obstacles for dislocation motion. 
Moreover, during indentation test additional vacancies are introduced into the affected region below the indent \cite{Gao2014} and these vacancies agglomerate with v+nH already existing in the material forming vacancy clusters. Hydrogen binding to vacancy clusters is even stronger than that for monovacancies~\cite{Myers1982,Maisonneuve2011}. Hence, vacancy clusters associated with hydrogen may represent efficient obstacles for motion of dislocations. 
Contrary to hydrogen bound to vacancies lose hydrogen dissolved in Nb lattice is known to increase the mobility of dislocations and to enhance the plasticity of material \cite{Robertson2001, Gahr1977}. Moreover, it has been reported that hydrogen dissolved in the lattice reduces the pop-in load in nanoindentation testing \cite{Barnoush2010}. This phenomenon has been attributed to reduced tension of dislocation line and/or reduced shear modulus in the presence of hydrogen. 
Hence, on the base of the available data, it is believed that formation of vacancies and their interaction with hydrogen is responsible for enhanced hardness in the sub-surface region of Nb samples studied.

Results of instrumented nanoindentation showed that mechanical properties of sub-surface region of the samples provided by Ningxia and Tokyo Denkai are similar and only very slightly affected by baking (changes in the characteristic length $h^{\star}$  are insignificant) despite the fact that VEPAS investigations showed at least partial release of hydrogen from vacancies and vacancy clusters by baking at $120^\mathrm{o}\,\mathrm{C}$. As one can see in Fig.~\ref{fig:ELBE_78} baking at $120^\mathrm{o}\,\mathrm{C}$ resulted in a decrease of the concentration of monovacancies but not vacancy clusters. It indicates that enhanced hardness  in the sub-surface region might be connected mainly with vacancy clusters. However, it is clear definite conclusion cannot be done at the present stage of knowledge and additional more detailed investigations are necessary to disclose the actual hardening  mechanism in the sub-surface region of Nb samples.   
In summary the ISE effect in all Nb samples studied cannot be explained only by considering the Nix-Gao model of monotonically increasing density of GNDs. Point defects, namely vacancies, vacancy clusters and hydrogen impurities seem to play important role in particular in the near-surface region.

\section{\label{sec:level4}Discussion}
A difference of the material supplied by Ningxia and Tokyo Denkai is obvious. Comparing the measurement of as-received samples, more vacancies in the bulk are observed for the Tokyo Denkai material, see Fig. \ref{fig:VacConc}. Another difference of the starting material is the local chemical environment of positron annihilation sites, which is visualized in the Fig.~\ref{fig:CBD}. The Ningxia material has a slightly higher amount of hydrogen bound to vacancies than the Tokyo Denkai material, which is in agreement with the chemical composition measured for the two ingots. The ingot from which the Ningxia material was cut showed a slightly higher content in C, N, O and H compared to the Tokyo Denkai ingot. Higher initial concentration of hydrogen in the sample 6 is confirmed by shorter positron lifetime $\tau_2$, see Fig. \ref{fig:Lifetime}. This is in accordance with the nanoindentation measurement shown in Fig. \ref{fig:Indent_before}. The higher hardness of sample 6 is likely a consequence of the higher amount of interstitials \cite{Ito}. The actual shape of the hardness curves for both samples is similar. This indicates that the density of dislocation motion obstacles is similar in both samples. Analysis of the ISE in both samples indicate that a significant density of vacancies or vacancy clusters is likely present in the near surface region. 
\newline
The results presented here can explain inconsistencies observed and discussed in \cite{Romanenko2013} and \cite{Visentin2010}. The samples in \textit{Visentin et al.}\cite{Visentin2010} were baked for about 2.5 - 4\,h at $120^\mathrm{o}\,\mathrm{C}$ and showed an increase in the DB-VEPAS S-parameter while the cavity cut-outs in \textit{Romanenko et al.} \cite{Romanenko2013} which showed less vacancies were baked for 48\,h at $120^\mathrm{o}\,\mathrm{C}$. As it has been shown here, these measurements are by all means in agreement, since the duration of the baking is an important parameter (see figure \ref{fig:VacConc}) and a 4\,h bake increases the vacancy concentration while a 48\,h or longer time bake leads to a reduced vacancy concentration. This emphasizes the importance of study of the dynamics of the vacancy-hydrogen interaction in Nb material for superconductive cavities during the baking procedure which has never been investigated so far and the results shown here represent the first attempt to identify potential significance of vacancy-hydrogen interaction for our understanding of Nb superconducting cavity performance.
\newline
It can be concluded from our bulk CDB measurements that hydride particles present in the as-received samples start to decompose and released hydrogen is trapped at vacancies resulting in formation of v+nH complexes, see figure \ref{fig:CBD}. The average number n of hydrogen atoms bound to vacancies in the bulk increases in the temperature range from $70^\mathrm{o}\,\mathrm{C}$ to $120^\mathrm{o}\,\mathrm{C}$. The v+nH complexes in the bulk are then stable at $120^\mathrm{o}\,\mathrm{C}$ for at least 4\,h. For longer baking times, the dissociation of these complexes in the bulk starts and the hydrogen is released back into the lattice. 
\newline
From the VEPALS measurements during in-situ baking of samples, the existence of v+nH complexes in the near surface region is observed as well, see Fig. \ref{fig:ELBE_78}. A high load of 4 or more hydrogen atoms has been observe. The baking procedure dissociated most hydrogen atoms but not fully released all the hydrogen from the vacancies. In addition, the existence of v+nH complexes was detected also  in cavity cut-outs, see Fig. \ref{fig:ELBE_CutOuts_three}. Whether the observed release in bulk and near the surface is a gradual or abrupt process has to be studied further but this could be an important feature.  
\newline
This is because the exact time scales for the hydrogen release are of importance importance. In \cite{Misiorek1993} it was argued, that in the case of large amount of unoccupied sites in the lattice one faces disordered structure when a chaotic distribution of rapidly moving hydrogen atoms appears, but if the occupation of these positions proceeds in an ordered way a more stable structure will emerge. Additionally, it has been found \cite{Birnbaum1976} that the growth of the $\beta$-NbH is diffusion controlled and that rapid hydrogen diffusion influences the phase transformation $\alpha \rightarrow \beta$-NbH and less ordered NbH phases form within $\beta$-NbH~\cite{Isagawa1980}. Another contribution is, that the H-H self-trapping energies increase with increasing hydrogen concentration, indicating that hydrogen diffusion is locally hindered with increasing hydrogen concentration \cite{Steiger, Ford2013b}. Hence, a moderated controlled release and diffusion can influence the formation of lossy nanohydrides. 
\newline
The  intermediate baking step at $75^\mathrm{o}\,\mathrm{C}$ for 4\,h might be moderating the hydrogen distribution in the lattice and therefore optimizing the formation of niobium phases and influencing the cavity performance. The co-existence of different phases in cavity cut-outs has been shown \cite{Trenikhina2015b}, where two phases of NbH, namely the less ordered $\epsilon$ and the more ordered $\beta$ phases, were observed at $\mathrm{LN_2}$ temperatures and the formation of two phases of NbH on certain samples has been reported~\cite{Barkov2013}. Interestingly, only the smaller hydrides of the two phases are formed in unbaked samples where cavities treated that way would normally show the high field Q-slope, hinting that they are the cause for the additional losses. This observation is well in agreement with the superconducting by proximity model mentioned above. An exponential dependency of the surface resistance of a cavity on the duration of the baking time has been observed \cite{Kneisel1999}, which in this model would agree with a gradual release of hydrogen and strengthens the need to investigate the dynamic behavior. In Ref.~\cite{Ciovati2004}, the hydrogen concentration depth profile in the first 100\,nm was measured using nuclear reaction analysis on samples with different annealing temperature showing a continuous reduction in the surface layer, correlating with an improvement of the surface resistance.    
\newline
It is interesting to connect the microstructure findings with the  observations made for the European XFEL cavity production which are still not explained. The average surface losses at 4\,MV/m of 234 cavities made from Ningxia material were significantly lower than those of 406 cavities made from Tokyo Denkai material \cite{Reschke2017} and the average maximum accelerating field was higher for Ningxia material than Tokyo Denkai material and no reason within the available fabrication and treatment data has been found for this since all specifications were met by both niobium vendors. Our measurements might be able to explain why on average Ningxia material should have a lower surface resistance than Tokyo Denkai. Less v+nH complexes remain in Tokyo Denkai material after the baking procedures compared to Ningxia material, which means more hydrogen was released into niobium for Tokyo Denkai material, leading to possible more nanohydrides forming and increasing the surface resistance. However, when comparing results for these samples with the XFEL cavity performance one has to take into account that, contrary to the annealing procedure applied in this work, the standard procedure applied during the European XFEL cavity production does not include the low temperature step at $75^\mathrm{o}\,\mathrm{C}$. Hence, if this lower T step influences the dynamics of  the interaction of hydrogen with vacancies, the result is not directly transferable. In addition, the measurements presented here are single measurements on two samples and more statistics is needed to exclude a falsification by an individual sheet.  
\newline
Another link to an actual cavity performance was found in the VEPALS measurements of samples cut from a cavity which was studied in detail. The results shown in Fig. \ref{fig:ELBE_CutOuts_Mean} fit well together with the observations from the cavity test. The difference in the lattice structure observed in the present work for sample 1, represented by the mean positron lifetime $\tau_{mean}$, explains the significant amount of trapped flux during the cavity test, limiting the cavity performance to a high degree, showing again the potential of the methods used in this paper to improve our understanding of fundamental processes happening during cavity surface treatments.
\newline
Another recent finding of the low-T bake, is a 'branching behavior' of the cavity \cite{Grassellino2019, Palczweksi2019} which is not understood yet. This branching behaviour shows that a cavity subjected to the low-T bake can be in one of two 'resistance states' with a difference of about 50\% in the achievable accelerating field and surface losses between those two states. The parameter which determines to which branch the cavity will fall  seems to be the starting temperature of the cool down, although it is currently investigated if this is the only parameter. The starting temperature is the equilibrium temperature of the cryostat at which a cavity is hold for several hours before cooling down. This branching behaviour can be explained by v+nH complexes as well. Using internal friction measurements it has been found~\cite{Yoshinari1982} that for an increase of the number of hydrogen atoms located at vacancies, the extent of the formation of hydrides and the temperature at which the formation takes place are shifted towards higher temperatures. Hence, if the starting temperature of the cryostat is close to the hydride formation temperature and kept there for several hours, possible disadvantageous phases could form but if the starting temperature is above the formation temperature and the cavity is cooled down quickly across the relevant temperature range, the formation of these phases is suppressed, hence two scenarios of surface losses are introduced. This effect plays a role, when v+nH complexes with 4 or more hydrogen atoms are present in the depth of 10-200\,nm below the surface. This is indeed indicated by the measurements presented here, such as of sample 78 (Fig.~\ref{fig:ELBE_78}) and the cavity cut-outs (Fig.~\ref{fig:ELBE_CutOuts_three}).

\section{\label{sec:level5}Conclusions}
Although vacancy-hydrogen interaction in the niobium lattice at the baking temperature and cryogenic temperatures have been the scope of research for quite some time, only with the results presented here, the dynamics of the v+nH complexes formation have been shown to be of high importance. Both, the baking temperature and the duration of baking plays a crucial role in the dynamics of absorbed hydrogen and vacancies. Our findings merged two previous contradicting results showing that the solely existence of vacancies and v+nH complexes in the rf layer might be not enough to explain the observed rf performance, but one has to consider also  the kinetics of hydrogen diffusion, trapping in vacancies and growth of hydride phases over time. 
\newline
Low temperature baking causes decomposition of hydride particles and released hydrogen is trapped at vacancies forming vacancy hydrogen complexes. As a consequence low temperature baking in the temperature range of $80-120^\mathrm{o}\,\mathrm{C}$ leads to an increase of the vacancy concentration. Prolonged annealing at $120^\mathrm{o}\,\mathrm{C}$ results in gradual removal of vacancy-hydrogen complexes in the bulk due to thermally activated release of hydrogen from vacancies. Remaining vacancies disappear by diffusion to sinks or agglomerate into small vacancy clusters. During prolonged baking at $120^\mathrm{o}\,\mathrm{C}$ vacancy clusters are gradually annealed out. 

In the sub-surface region vacancies are agglomerated into vacancy clusters  decorated by hydrogen  and low temperature baking at $70^\mathrm{o}\,\mathrm{C}$ leads to detachment of hydrogen from the vacancy complexes, whereas annealing at $120^\mathrm{o}\,\mathrm{C}$ significantly reduces the number of v+nH complexes.

\section{\label{sec:level6}Acknowledgments}
The authors of this work would like to thank A. Ermakov and S. Sievers (DESY) for their support in the preparation of the samples and R. Krause-Rehberg (TU Halle) for the discussion of the results. Positron annihilation lifetime studies were carried out at the ELBE instrument at the Helmholtz-Zentrum Dresden-Rossendorf.

This work was supported by the BMBF under the research grant 05H18GURB1, the Impulse-und Networking fund of the Helmholtz Association (FKZ VH-VI-442 Memriox), and the Helmholtz Energy Materials Characterization Platform (03ET7015). The indentation was carried out in the frame of the project CZ.02.1.01/0.0/0.0/16\_019/0000778 (Centre of Advanced Applied Sciences).

\section{Author contributions}
M.W., M.O.L. and J.C. wrote the main manuscript and all authors reviewed the manuscript. M.W. and C.B. prepared the samples. M.W. and H.W. designed the research and M.W. performed a comparison with existing literature. M.W., C.B. M.O.L. did the PALS measurements and M.B., M.O.L.,E.H. and A.W. prepared the PAS and PALS experiments. M.O.L. did the PAS measurements and analyis while M.B. and M.O.L. did the PALS analysis. P.H. did the nanoindent measurements and J.C. did the CDB measurements.
Fig. 1, 3, 8, 9, 10, 11, 12, 13 were prepared by J.C.. Fig. 4, 5, 6, 7  were prepared by M.O.L and M.B.. Fig. 2 was prepared by M.W..

\section{Competing interests}
The authors declare no competing interests.


\begin{thebibliography}{99}

\bibitem{Bonin1991} Bonin, B. \& R{\"{o}}th, R., Q Degradation of Niobium Cavities due to Hydrogen Contamination \textit{Particle Accelerators} \textbf{40}, 59-83 (1992).

\bibitem{Antoine2003} Antoine, C. \& Berry, S. H in Niobium: Origin And Method Of Detection. \textit{AIP Conf. Proc.} \textbf{671}, 176 (2003).

\bibitem{Isagawa1980} Isagawa, S. Hydrogen absorption and its effect on low-temperature electric properties of niobium. \textit{J. Appl. Phys.} \textbf{51}, 4460-4470 (1980).

\bibitem{Aderhold2010} Aderhold, S. et~al. Cavity Process. \textit{ILC HiGrade Reports} \url{http://www.ilc-higrade.eu/e83212/e99561/e99569/ILC-HiGrade-2010-005-1.pdf} (2015).

\bibitem{Ciovati2010a} Ciovati, G., Myneni, G., Stevie, F., Maheshwari, P. \& Griffis, D. High field Q slope and the baking effect: Review of recent experimental results and new data on Nb heat treatments. \textit{Phys. Rev. Accel. Beams} \textbf{13}, 22002 (2010).

\bibitem{Cottrell1949} Cottrell, A.~H. \& Bilby, B. Dislocation Theory of Yielding and Strain Aging of Iron. \textit{Proceedings of the Physical Society. Section A} \textbf{62}, 49-62 (1949).

\bibitem{Chemical2006} Khaldeev, G.~V. \& Gogel, V.~K. Physical and Corrosion-electrochemical Properties of the Niobium–Hydrogen System. \textit{Russ. Chem. Rev.} \textbf{56}, 605-618 (1987).

\bibitem{Cizek2009} Cizek, J. et~al. Hydrogen-vacancy complexes in electron-irradiated niobium. \textit{Phys. Rev. B} \textbf{79}, 054108 (2009).

\bibitem{Romanenko2010} Romanenko, A. \& Padamsee, H. The role of near-surface dislocations in the high magnetic field performance of superconducting niobium cavities. \textit{Supercond. Sci. Technol} \textbf{23}, 45008 (2010).

\bibitem{Romanenko2013} Romanenko, A., Edwardson, C.~J., Coleman, P.~G. \& Simpson, P.~J. The effect of vacancies on the microwave surface resistance of niobium revealed by positron annihilation spectroscopy. \textit{Appl. Phys. Lett.} \textbf{102}, 232601 (2013).

\bibitem{Ford2013b} Ford, D.~C. et~al. First-principles calculations of niobium hydride formation in superconducting radio-frequency cavities. \textit{Supercond. Sci. Technol} \textbf{26}, 95002-9 (2013).

\bibitem{DeGennes1965} De Gennes, P.~G. \& Hurault, J.~P. Proximity effects under magnetic fields. II - Interpretation of breakdown. \textit{Physics Letters} \textbf{17}, 181-182 (1965).

\bibitem{Fauchere1997} Fauch{\`{e}}re, A.~L. \& Blatter, G. Magnetic breakdown in a normal-metal–superconductor proximity sandwich. \textit{Phys. Rev. B} \textbf{56}, 14102-14107 (1997).

\bibitem{Visentin2010} Visentin, B., Barthe, M.~F., Moineau, V. \& Desgardin, P. Involvement of hydrogen-vacancy complexes in the baking effect of niobium cavities. \textit{Phys. Rev. Accel. Beams} \textbf{13}, 052002 (2010).

\bibitem{Ciovati2007a} Ciovati, G., Kneisel, P. \& Gurevich, A. Measurement of the high-field Q drop in a high-purity large-grain niobium cavity for different oxidation processes. \textit{Phys. Rev. Accel. Beams} \textbf{10}, 1-19 (2007).

\bibitem{Romanenko2013b} Romanenko, A., Barkov, F., Cooley, L.~D. \& Grassellino, A. Proximity breakdown of hydrides in superconducting niobium cavities. \textit{Supercond. Sci. Technol} \textbf{26}, 035003 (2013).

\bibitem{Trenikhina2015c} Trenikhina, Y., Romanenko, A., Kwon, J., Zuo, J.-M. \& Zasadzinski, J.~F. Nanostructural features degrading the performance of superconducting radio frequency niobium cavities revealed by transmission electron microscopy and electron energy loss spectroscopy. \textit{J. Appl. Phys.} \textbf{117}, 164904 (2015).

\bibitem{Grassellino2018} Grassellino, A. et~al. Accelerating fields up to 49 MV/m in TESLA-shape superconducting RF niobium cavities via 75C vacuum bake. Preprint at \url{http://arxiv.org/abs/1806.09824} (2018).
 
\bibitem{Stanley1967} Stanley, M.~W. \& Szkopiak, Z.~C. The alpha and beta Peaks in Cold-Worked Niobium. \textit{J. Mater. Sci.} \textbf{2}, 559-566 (1967).

\bibitem{Gupta1994} Gupta, C.~K. \& Suri, A.~K. in \textit{Extractive Metallurgy of Niobium} (CRC Press, 1994).

\bibitem{Barkov2012} Barkov, F., Romanenko, A. \& Grassellino, A. Direct observation of hydrides formation in cavity-grade niobium. \textit{Phys. Rev. Accel. Beams} \textbf{15}, 122001 (2012).

\bibitem{Barkov2013} Barkov, F., Romanenko, A., Trenikhina, Y. \& Grassellino, A. Precipitation of hydrides in high purity niobium after different treatments. \textit{J. Appl. Phys.} \textbf{114}, 164904 (2013).

\bibitem{Richter1976} Richter, D., T{\"{o}}pfler, J. \& Springer, T. The influence of dissolved nitrogen on hydrogen diffusion in niobium studied by neutron spectroscopy. \textit{J. Phys. Condens. Matter} \textbf{6}, L93 (1976).

\bibitem{Pfeiffer1976} Pfeiffer, G.\& Wipf, H. The trapping of hydrogen in niobium by nitrogen interstitials. \textit{J. Phys. Condens. Matter} \textbf{6}, 167-179 (1976).

\bibitem{Grassellino2013a} Grassellino, A. et~al. Nitrogen and argon doping of niobium for superconducting radio frequency cavities: a pathway to highly efficient accelerating structures. \textit{Supercond. Sci. Technol.} \textbf{26}, 102001 (2013).

\bibitem{Grassellino2017d} Grassellino, A. et~al. Unprecedented quality factors at accelerating gradients up to 45MVm-1 in niobium superconducting resonators via low   temperature nitrogen infusion. \textit{Supercond. Sci. Technol.} \textbf{30}, 094004 (2017).

\bibitem{Sung2019} Sung, Z., Romanenko, A.\& Grassellino, A. Niobium Hydride Studies using Cryo-AFM. \textit{Proceedings of TTC Workshop 2019}
\url{https://indico.desy.de/indico/event/21337/session/12/contribution/42/material/slides/0.pptx} (2019).


\bibitem{Brinkmann2014a} Brinkmann, R., Schneidmiller, E.~A., Sekutowicz, J.\& Yurkov, M.~V. Prospects for CW and LP operation of the European XFEL in hard X-ray regime. \textit{Nucl. Instrum. Methods Phys. Res A} \textbf{768},20-25 (2014).

\bibitem{Michizono2017} Evans, L. \& Michizono, S. The International Linear Collider Machine Staging Report 2017. Preprint at \url{http://arxiv.org/abs/1711.00568} (2017).
 
\bibitem{Raubenheimer2018a} Raubenheimer, T.~O. The LCLS-II-HE, A High Energy Upgrade of the LCLS-II. \textit{60th ICFA Advanced Beam Dynamics Workshop on Future Light Sources}, 6-11 (2018).

\bibitem{Zhu2018} Zhu, Z.et~al. SCLF: An 8-GeV CW SCRF Linac-Based X-Ray FEL Facility in Shangahi. \textit{38th International Free Electron Laser Conference}, 182-184 (2018).

\bibitem{Fukai1994} Fukai, Y.\& Okuma, N. Formation of Superabundant Vacancies in Pd Hydride under High Hydrogen Pressures. \textit{PRL} \textbf{73}, 1640-1643 (1994).

\bibitem{Altarelli2007} Altarelli, M.et~al. The Technical Design Report of the European XFEL. \textit{Technical Report} \url{http://xfel.desy.de/localfsExplorer{\_}read?currentPath=/afs/desy.de/group/xfel/wof/EPT/TDR/XFEL-TDR-final.pdf} (2006).
  
\bibitem{Singer2016} Singer, W. et~al. Production of superconducting 1.3-GHz cavities for the European X-ray Free Electron Laser. \textit{Phys. Rev. Accel. Beams} \textbf{19}, 092001 (2016).

\bibitem{Reschke2017} Reschke, D. et~al. Performance in the vertical test of the 832 nine-cell 1.3 GHz cavities for the European X-ray Free Electron Laser. \textit{Phys. Rev. Accel. Beams} \textbf{20}, 042004 (2017).

\bibitem{Hausild2016} Hau{\v{s}}ild, P., Materna, A., Kocmanov{\'{a},L. \&  Mat{\v{e}}j{\'{i}}{\v{c}}ek,J. Determination of the individual phase properties from the measured grid indentation data.  \textit{J. Mater. Res. Technol} \textbf{31}, 3538-3548 (2016).

\bibitem{Hausild2018} Hau{\v{s}}ild, P.,\v{C}}ech, J., Materna, A.\& Mat{\v{e}}j{\'{i}}{\v{c}}ek, J. Statistical treatment of grid indentation considering the effect of the interface and the microstructural length scale. \textit{Mech. Mater} \textbf{129}, 99-103 (2019).

\bibitem{Cizek2018} {\v{C}}{\'{i}}{\v{z}}ek, J. Characterization of lattice defects in metallic materials by positron annihilation spectroscopy: A review. \textit{J. Mater. Sci. Technol.} \textbf{34}, 577-598 (2018).

\bibitem{Krause-Rehberg1999} Krause-Rehberg, R.\& Leipner, H.~S.	\textit{Positron annihilation in semiconductors: defect studies} (Springer-Verlag, 1999).

\bibitem{West1973} West, R.~N. Positron studies of condensed matter. \textit{Adv. Phys.} \textbf{22}, 263-383 (1973).

\bibitem{Cizek2004} Cizek, J.et~al. Hydrogen-induced defects in bulk niobium. \textit{Phys. Rev. B} \textbf{69}, 224106 (2004).

\bibitem{Lynn1977} Lynn, K.~G.et~al. Positron-Annihilation Momentum Profiles in Aluminum: Core Contribution and the Independent-Particle Model. \textit{PRL} \textbf{38}, 241-244 (1977).

\bibitem{Brusa2002} Brusa, R.~S.,Deng, W., Karwasz, G.~P.\& Zecca, A. Doppler-broadening measurements of positron annihilation with high-momentum electrons in pure elements. \textit{Nucl. Instrum. Methods Phys. Res B} \textbf{194}, 519-531 (2002).

\bibitem{Schultz1988} Schultz, P.~J.\& Lynn, K.~G. Interaction of positron beams with surfaces, thin films, and interfaces. \textit{Rev. Mod. Phys.} \textbf{60}, 701 (1988).

\bibitem{Anwand2012} Anwand, W.,Brauer, G., Butterling, M.,Kissener, H.~R.\& Wagner, A. Design and construction of a slow positron beam for solid and surface investigations. \textit{Defect Diffus. Forum} \textbf{331}, 25-40 (2012).

\bibitem{Liedke2015} Liedke, M.~O.et~al. Open volume defects and magnetic phase transition in Fe60 Al40 transition metal aluminide. \textit{J. Appl. Phys.} \textbf{117}, 163908 (2015).

\bibitem{Wagner2018} Wagner, A., Butterling, M., Liedke, M.~O.,Potzger, K.\& Krause-Rehberg, R. Positron annihilation lifetime and Doppler broadening spectroscopy at the ELBE facility. In \textit{AIP Conf. Proc.}, 40003 (2018).

\bibitem{Wagner2017a} Wagner, A. et~al. Positron annihilation lifetime spectroscopy at a superconducting electron accelerator. \textit{J. Phys. Conf. Ser.} \textbf{791}, 012004 (2017).

\bibitem{Becvar2005} Be{\v{c}}v{\'{a}}{\v{r}}, F., {\v{C}}{\'{i}}{\v{z}}ek, J., Proch{\'{a}}zka, I.\& Janotov{\'{a}}, J. The asset of ultra-fast digitizers for positron-lifetime spectroscopy. \textit{Nucl. Instrum. Methods Phys. Res A} \textbf{539}, 372-385 (2005).

\bibitem{Cizek2010} {\v{C}}{\'{i}}{\v{z}}ek, J., Vl{\v{c}}ek, M. \& Proch{\'{a}}zka, I. Digital spectrometer for coincidence measurement of Doppler broadening of positron annihilation radiation. \textit{Nucl. Instrum. Methods Phys. Res A} \textbf{623}, 982-994 (2010).

\bibitem{Cizek2012} {\v{C}}{\'{i}}{\v{z}}ek, J., Vl{\v{c}}ek, M. \& Proch{\'{a}}zka, I. Investigation of positron annihilation-in-flight using a digital coincidence Doppler broadening spectrometer. \textit{New J. Phys.} \textbf{14}, 18 (2012).

\bibitem{PALSfit3} PALSfit3, Version 3.218 \url{http://palsfit.dk/}. 

\bibitem{OriginLab} Origin(Pro), Version 2019b, OriginLab Corporation, Northampton, MA, USA.

\bibitem{Becvar2007} Be{\v{c}}v{\'{a}}{\v{r}}, F., Methodology of positron lifetime spectroscopy: Present status and perspectives. \textit{Nucl. Instrum. Methods Phys. Res B} \textbf{261}, 871-874 (2007). 

\bibitem{Prochazka1997} Proch{\'{a}}zka, I., Novotn{\'{y}}, I. \& Be{\v{c}}v{\'{a}}{\v{r}}, F. Application of Maximum-Likelihood Method to Decomposition of Positron-Lifetime Spectra to Finite Number of Components. \textit{Mater. Sci. Forum} \textbf{255-257}, 772-774 (1997).

\bibitem{SigmaPlot} SigmaPlot, Version 14.0, Systat Software, San Jose, CA, USA.

\bibitem{Hugenschmidt2016} Hugenschmidt, C. Positrons in surface physics. \textit{Surf. Sci. Rep.} \textbf{71}, 547-594 (2016).

\bibitem{Chu1981} Chu, S., Mills, A.~P.\& Murray, C.~A. Thermodynamics of positronium thermal desorption from surfaces. \textit{Phys. Rev. B} \textbf{23}, 2060-2064 (1981).

\bibitem{Wenskat2019b} Wenskat, M. et~al. Cavity Cut-Out Studies of a 1.3 GHz Single-Cell Cavity after a failed Nitrogen Infusion Process. \textit{19th International Conference on RF Superconductivity} \url{http://accelconf.web.cern.ch/AccelConf/srf2019/papers/mop025.pdf} (2019). 

\bibitem{Antoine2019a} Antoine, C. Influence of crystalline structure on rf dissipation in superconducting niobium. \textit{Phys. Rev. Accel. Beams} \textbf{22}, 034801 (2019).

\bibitem{Nye53} Nye, J.~F. Some geometrical relations in dislocated crystals. \textit{Acta Mater.} \textbf{1}, 153-162 (1953).

\bibitem{Milman11} Milman, Y.~V.,Golubenko, A.~A.\& Dub, S.~N. Indentation size effect in nanohardness. \textit{Acta Mater.} \textbf{59}, 740-748 (2011).

\bibitem{Dub2010} Dub, S.~N., Lim, Y.~Y. \& Chaudhri, M.~M. Nanohardness of high purity Cu (111) single crystals: The effect of indenter load and prior plastic sample strain. \textit{J. Appl. Phys.} \textbf{107}, 43510 (2010).

\bibitem{Nix1998} Nix, W.~D. \& Gao, H. Indentation size effects in crystalline materials: A law for strain gradient plasticity. \textit{J. Mech. Phys. Solids.} \textbf{46}, 411-425 (1998).

\bibitem{Gerberich02} Gerberich, W.~W.,Tymiak, N.~I., Grunlan, J.~C.,Horstemeyer, M.~F.\& Baskes, M.~I. Interpretations of indentation size effects. \textit{J. Appl. Mech.} \textbf{69}, 433-442 (2002).

\bibitem{Manika06} Manika, I. \& Maniks, J. Size effects in micro- and nanoscale indentation. \textit{Acta Mater.} \textbf{54}, 2049-2056 (2006).

\bibitem{Kasada2011} Kasada, R., Takayama, Y., Yabuuchi, K., Kimura, A.  
A new approach to evaluate irradiation hardening of ion-irradiated ferritic alloys by nano-indentation techniques. \textit{Fusion Eng. Des.} \textbf{86}, 2658-2661 (2011).

\bibitem{Kasada2014} Kasada, R., Konishi, S., Yabuuchi, K., Nogami, S., Ando, M., Hamaguchi, D., Tanigawa, H.
Depth-dependent nanoindentation hardness of reduced-activation ferritic steels after MeV Fe-ion irradiation. \textit{Fusion Eng. Des.} \textbf{89}, 1637-1641 (2014).

\bibitem{Kasada2016} Kasada, R., Konishi, S., Hamaguchi, D., Ando, M., Tanigawa, H.
Evaluation of strain-rate sensitivity of ion-irradiated austenitic steel using strain-rate jump nanoindentation tests. \textit{Fusion Eng. Des.} \textbf{109-111}, 1507-1510 (2016).

\bibitem{Hasenhuetl2017} Hasenhuetl, E., Kasada, R., Zhang, Z., Yabuuchi, K., Huang, Y.-J., Kimura, A. 
Evaluation of Ion-Irradiation Hardening of Tungsten Single Crystals
by Nanoindentation Technique Considering Material Pile-Up Effect. \textit{Materials Transactions} \textbf{58}, 749-756 (2017).

\bibitem{Ruiz-Moreno2018} Ruiz-Moreno, A., H\" ahner, P. 
Indentation size effects of ferritic/martensitic steels: A comparative
experimental and modelling study. \textit{Mater. Des.} \textbf{145}, 168-180 (2018).

\bibitem{Huang2006} Huang, Y., Zhang, F., Hwang, K.C., Nix, W.D., Pharr G.M., Feng, G. 
A model of size effects in nano-indentation. \textit{J. Mech. Phys. Sol.} \textbf{54}, 1668-1686 (2006).

\bibitem{Myers1982} Myers, S.M.,  Follstaedt, D.M., Besenbacher, F., B\o ttiger, J. 
Trapping and surface permeation of
deuterium in He-implanted Fe. \textit{J. Appl. Phys.} \textbf{53}, 8734-8744 (1982).

\bibitem{Maisonneuve2011} Maisonneuve, J., Oda, T., Tanaka, S. 
Molecular Statics Study of Hydrogen Isotope
Trapping in BCC-Iron Vacancy Clusters. \textit{Fus. Sci. Technol.} \textbf{60}, 1507-1510 (2011).

\bibitem{deGuzman93} Shell-De-Guzman, M., Neubauber, G., Flinn, P.\& Nix, W.~D. Role of indentation depth on the measured hardness of materials. \textit{Materials Research Society Symposium - Proceedings} \textbf{308}, 613-618 (1993).

\bibitem{Li2015} Li, S.et~al. The interaction of dislocations and hydrogen-vacancy complexes and its importance for deformation-induced proto nano-voids formation in $\alpha$-Fe. \textit{Int. J. Plast.} \textbf{74}, 175-191 (2015).

\bibitem{Gao2014} Gao, Y.,Ruestes, C.~J.\& Urbassek, H.~M. Nanoindentation and nanoscratching of iron: Atomistic simulation of dislocation generation and reactions. \textit{Comput. Mater. Sci.} \textbf{90}, 232-240 (2014).

\bibitem{Robertson2001} Robertson, I.~M. The effect of hydrogen on dislocation dynamics. \textit{EFM} \textbf{68}, 671-692 (2001).

\bibitem{Gahr1977} Gahr, S.,Grossbeck, M.~L.\& Birnbaum, H.~K. Hydrogen embrittlement of Nb I-Macroscopic behavior at low temperatures. \textit{Acta Mater.} \textbf{25}, 125-134 (1977).

\bibitem{Barnoush2010} Barnoush, A.\& Vehoff, H. Recent developments in the study of hydrogen embrittlement: Hydrogen effect on dislocation nucleation. \textit{Acta Mater.} \textbf{58}, 5274-5285 (2010).

\bibitem{Zhou2016} Zhou, X.,Ouyang, B., Curtin, W.~A.\& Song, J. Atomistic investigation of the influence of hydrogen on dislocation nucleation during nanoindentation in Ni and Pd. \textit{Acta Mater.} \textbf{116}, 364-369 (2016).

\bibitem{Ito} Ito, M. Studies on Physical and Hydrogen Properties Behavior of Metal Hydrides in Zr Alloys. Ph.D. thesis, Osaka University (2008).
 
\bibitem{Misiorek1993} Misiorek, H.,Jezowski, A., Mucha, J.\& Sorokina, N.~I. Hysteresis of thermal conductivity and electrical resistivity of niobium hydrides. \textit{Solid. State. Commun.} \textbf{85}, 907-910 (1993).

\bibitem{Birnbaum1976} Birnbaum, H.,Grossbeck, M.\& Amano, M. Hydride Precipitation in Nb and some Properties of NbH. \textit{J. Less. Common. Met.} \textbf{49}, 357-370 (1976).

\bibitem{Steiger} Steiger, J.,Blasser, S.\& Weidinger, A. Solubility of hydrogen in thin niobium films. \textit{Phys. Rev. B} \textbf{49}, 5570-5574 (1994).

\bibitem{Trenikhina2015b} Trenikhina, Y.\& Romanenko, A. Nanostructure of the Penetration Depth in Nb Cavities: Debunking the Myths and New Findings. \textit{19th International Conference on RF Superconductivity} \url{http://accelconf.web.cern.ch/AccelConf/SRF2015/papers/wea1a05.pdf} (2015).

\bibitem{Kneisel1999} Kneisel, P. Preliminary Experience with 'in-Situ' Baking of Niobium Cavities. \textit{9th International Conference on RF Superconductivity} \url{http://accelconf.web.cern.ch/AccelConf/SRF99/papers/tup044.pdf} (1999).

\bibitem{Ciovati2004} Ciovati, G. Effect of low-temperature baking on the radio-frequency properties of niobium superconducting cavities for particle accelerators. \textit{J. Appl. Phys.} \textbf{96}, 1591-1600 (2004).

\bibitem{Grassellino2019} Grassellino, A. Progress in high Q and high gradient. \textit{Proceedings of TTC Workshop 2019}
\url{https://indico.desy.de/indico/event/21337/session/9/contribution/5/material/slides/0.pdf} (2019).

\bibitem{Palczweksi2019} Palczewski, A. High Q0/High gradient at JLab: LCLS-2 HE 3N6 doping, furnace issues and FNAL 75C retests. \textit{Proceedings of TTC Workshop 2019} \url{https://indico.desy.de/indico/event/21337/session/14/contribution/48/material/slides/0.pptx} (2019).
  
\bibitem{Yoshinari1982} Yoshinari, O.\& Koiwa, M. Low Frequency Internal Friction Study of V-H, Bb-H and Ta-H Alloys. \textit{Acta Mater.} \textbf{30}, 1979-1986 (1982).

\end{thebibliography}
\end{document}